# DETECTION OF NE VIII IN THE LOW-REDSHIFT WARM-HOT IGM[1]

Blair D. Savage[2], Nicolas Lehner[2], Bart P. Wakker[2],

Kenneth R. Sembach[3] and Todd M. Tripp[4]

## ABSTRACT

High resolution FUSE and STIS observations of the bright QSO HE 0226-4110 ($z_{em}$ = 0.495) reveal the presence of a multi-phase absorption line system at $z_{abs}$(O VI) = 0.20701 containing absorption from H I (Ly α to Ly θ), C III, O III, O IV, O VI, N III, Ne VIII, Si III, S VI and possibly S V. Single component fits to the Ne VIII and O VI absorption doublets yield logN(Ne VIII) = 13.89±0.11, b =23±15 km s⁻¹, v = -7±6 km s⁻¹, and logN(O VI) = 14.37±0.03, b =31±2 km s⁻¹, v = 0±2 km s⁻¹. The Ne VIII and O VI doublets are detected at 3.9σ and 16σ significance levels, respectively. This represents the first detection of intergalactic Ne VIII, a diagnostic of gas with temperature in the range from ~5x10⁵ to ~1x10⁶ K. Through the entire absorber, N(Ne VIII)/N(O VI) = 0.33±0.10. The O VI and Ne VIII are not likely created in a low density medium photoionized solely by the extragalactic background at z = 0.2 since the required path length ~11 Mpc (assuming [Z/H] = -0.5) implies the Hubble flow absorption line broadening would be ~10 times greater than the observed line widths. A collisional ionization origin is therefore more likely. Assuming [Ne/H] and [O/H] = -0.5, the value N(Ne VIII)/N(O VI) = 0.33±0.10 is consistent with gas in collisional ionization equilibrium near T=5.4x10⁵ K with logN(H)= 19.9 and N(H)/N(H I) = 1.7x10⁶. Various non-equilibrium ionization processes are also considered because gas with T~ (1-6)x10⁵ K cools efficiently. The observations of O VI and Ne VIII in the z = 0.20701 system support the basic idea that a substantial fraction of the baryonic matter at low redshift exists in hot very highly ionized gaseous structures. Absorption by the moderately ionized gas (including O IV and S VI) is well modeled by gas in photoionization equilibrium with [Z/H] = -0.5±0.2, logU ~ -1.85, T ~ 2.1x10⁴ K, $n_H$~2.6x10⁻⁵ cm⁻³, P/k ~ 0.5 cm⁻³ K, and path length of ~60 kpc. These values suggest the moderately ionized absorber may be associated with the modestly enriched photoionized gas in a galaxy group or the outermost regions of a galaxy halo.

*Subject Headings:* intergalactic medium – quasars: absorption lines – quasars: general – quasars:individual (HE 0226 –4110) – cosmology: observations

---

[1] Based partly on observations with (1) the NASA/ESA Hubble Space telescope obtained at the Space Telescope Science Institute, which is operated by the Association of Universities for Research in Astronomy, Inc., under NASA contract NAS5-26555, and (2) the NASA-CNES/ESA Far-ultraviolet Spectroscopic Explorer operated by Johns Hopkins University, supported by NASA contract NAS5-32985.

[2] Department of Astronomy, University of Wisconsin-Madison, 475 N. Charter St., Madison, WI 53706, USA.

[3] Space Telescope Science Institute, 3700 San Martin Drive, Baltimore, MD 21218, USA.

[4] Department of Astronomy, University of Massachusetts, Amherst, MA 01003, USA.

# 1. INTRODUCTION

Shock heated gas in the warm-hot intergalactic medium (WHIM) with temperatures ranging from $10^5$ to $10^7$ K may contain a substantial fraction of the baryons in the Universe at low redshift (Cen & Ostriker 1999; Davé et al. 2001). The warm component of the WHIM with T between $10^5$ and $10^6$ K has been revealed by the high frequency of occurrence of O VI λλ1031.926, 1037.617 absorption in QSO absorption line systems (Tripp, Savage & Jenkins 2000; Savage et al. 2002) and by the existence of a population of very broad Lyman α lines (Richter et al. 2004; Sembach et al. 2004). Detection of the hot component of the WHIM with T > $10^6$ K is possible through measurements of O VII λ21.602 and O VIII λ18.970 absorption at X-ray wavelengths. A WHIM absorber near z = 0 may have been detected with the Chandra Observatory (Nicastro et al. 2002) and XMM-Newton (Rasmussen, Kahn, & Paerels 2003). However, the origin of the z = 0 X-ray absorption is confused with foreground absorption produced by a hot corona to the Milky Way and other local structures (Fang et al. 2003; Futamoto et al. 2003; McKernan et al. 2004; Savage et al. 2005; Yao & Wang 2005). The early claims of O VII and O VIII absorption at low redshift beyond the local group (Fang et al. 2002; Mathur et al. 2003), are either of marginal significance or are contradicted by more recent observations with higher signal-to-noise ratios (S/N) (Rasmussen et al. 2003). However, the recent flaring of the Blazar Mrk 421 has provided a high S/N X-ray spectrum of two intervening systems containing O VII but no O VI (Nicastro et al. 2005). These two systems are evidently tracing the WHIM at temperatures from ~$10^{5.8}$ to $10^{6.4}$ K (Nicastro et al. 2005). It is now clear the WHIM can be effectively studied by looking for absorption by highly ionized atoms at X-ray, ultraviolet (UV), far-UV (FUV) and extreme-UV (EUV) wavelengths or through the detection of very broad Lyman α absorbers.

Measuring the physical conditions in the low redshift absorbers tracing highly ionized gas including O VI and other ions is an important first step toward determining the baryonic content of these absorbers. The baryonic content estimate will depend on the ionization conditions, the metallicity, and the frequency of occurrence of the different absorber types. Low redshift absorbers containing O VI have been found to have a wide range of properties. Some appear to be photoionized by the extragalactic background radiation (Savage et al. 2002). In others the ionization is probably from electron collisions in hot gas (Tripp et al. 2001). The O VI absorbers often trace multiphase media containing warm photoionized gas and hot collisionally ionized gas (Tripp et al. 2000, 2002; Shull et al. 2003; Richter et al. 2004; Sembach et al. 2004).

The physical properties and ionization conditions in IGM absorbers containing O VI can be most effectively studied by obtaining observations with broad spectroscopic wavelength coverage. EUV lines are important because they make possible the study of a wide range of ion types and provide access to tracers of gas at high temperature. For example, EUV measures of O II, O III, O IV, and O V can be combined with UV and FUV measures of O I and O VI to study the relationships among six different ions of oxygen. Similarly, UV to EUV observations allow studies of N I, N II, N III, N IV, and N V. Having observations of multiple ionization states of the same element greatly enhances understanding the origin(s) of the ionization in the gas, since the results are relatively insensitive to the gas metallicity. Access to a wide range of ions makes it easier to probe the multiphase nature of the absorbers.

The EUV lines of Ne VIII $\lambda\lambda$ 770.409, 780.324 and Mg X $\lambda\lambda$ 609.793, 624.941 are very important tracers of hot gas. In collisional ionization equilibrium (CIE), these two species peak in temperature at $7x10^5$ K and $1.1x10^6$ K, respectively (Sutherland & Dopita 1993). Ne VIII is particularly important for probing hot collisionally ionized gas because of the relatively high cosmic abundance of Ne and the relatively large f-values for the two Li-like resonance lines ($f_{770} = 0.103$, $f_{780} = 0.0505$; Verner, Verner & Ferland 1996). In addition, only modest redshifts (z > 0.18) are required to shift the Ne VIII lines into the wavelength range currently accessible to the Far Ultraviolet Spectroscopic Explorer (FUSE) satellite.

In hot gas with a solar abundance ratio of neon to oxygen in CIE, the ratio of Ne VIII to O VI exceeds 1.0 for T > $6x10^5$ K except for the narrow temperature range between 1.4 to $2.6x10^6$ where Ne VIII/ O VI decreases to 0.6 (see §5). Therefore, hot gas absorbing systems containing O VI are expected to contain detectable amounts of Ne VIII so long as T exceeds ~$6x10^5$ K. While there have been tantalizing hints of Ne VIII absorption in several O VI systems (see Richter et al. 2004), a clear detection is reported for the first time in this paper where we analyze high quality FUSE and Space Telescope Imaging Spectrograph (STIS) observations of the bright QSO HE 0226-4110 ($z_{em} = 0.495$). The FUSE observations cover the wavelength region from 916 to 1188 Å with a spectral resolution of ~20 km s$^{-1}$. The STIS observations extend from 1160 to 1708 Å with a resolution of 7 km s$^{-1}$. The observations reveal a multiphase O VI absorption line system at z = 0.20701 containing H I (Ly $\alpha$ to Ly $\theta$), C III, O III, O IV, O VI, N III, Ne VIII, Si III, and S VI. We use the observations to determine the ionization conditions and abundances in this multiphase absorber.

The organization of this paper is as follows: The observations are described in §2. The observed properties of the absorber are discussed in §3. The origin and abundances of the moderately ionized gas absorption containing the ions C III, O III, N III, and Si III are considered in §4. We find that most of the S VI and O IV we detect can be produced by photoionization in this moderately ionized gas. A photoionization origin for the O VI and Ne VIII is ruled out because Hubble flow broadening would be ~10x larger than the observed line widths over the large required path length (see §5). The highly ionized gas absorption containing O VI and Ne VIII is shown to occur in hot collisionally ionzed gas (see §5). The results are summarized in §6.

## 2. OBSERVATIONS

We have combined high resolution spectroscopic observations of HE 0226-4110 from STIS and FUSE to obtain measurements with nearly continuous wavelength coverage from 916 to 1707 Å. With a flux ranging from $3x10^{-14}$ to $2x10^{-14}$ to $1.5x10^{-14}$ erg cm$^{-2}$ s$^{-1}$ Å$^{-1}$ from 930 to 1400 to 1700 Å, HE 0226-411 is one of the few moderate redshift ($z_{em} = 0.495$) QSOs bright enough to obtain measurements with moderate S/N (~10 to ~20 per resolution element) in the UV and far-UV. All velocities in this paper are given in the heliocentric reference frame. Note that in the direction to HE 0226-4110 (l = 253.94°, b = -65.77°) the Local Standard of Rest (LSR) and heliocentric reference frames are related by $v_{LSR} = v_{HELIO} - 14.3$ km s$^{-1}$.

## 2.1. STIS Observations

The STIS observations of HE 0226-4110 were obtained over the time period 2002-Dec-26 to 2003-Jan-01 as part of a large survey to study highly-ionized intergalactic gas at low redshift (Guest Observer program GO-9184, see Table 1). Information about STIS and its inflight performance is given by Woodgate et al. (1998), Kimble et al. (1998), and in the STIS Instrument Handbook (Proffitt et al. 2002). We used the intermediate resolution echelle far-UV grating (E140M) and the 0.2"x0.06" entrance slit and obtained a total integration time of 43.5 ksec. The measurements cover the wavelengths between 1144 and 1709 Å with small gaps between the echelle orders for $\lambda > 1630$ Å. The spectral resolution (FWHM) is 7 km s$^{-1}$ with a detector pixel size of 3.5 km s$^{-1}$.

Our data reduction steps are the same as those described in our earlier IGM investigations with STIS (see Tripp et al. 2001 and Savage et al. 2002) and continued in the more recent investigations by Richter et al. (2004) and Sembach et al. (2004). The reduction includes a two dimensional scattered light correction and the weighted co-addition of individual extracted spectra to produce the final composite spectrum on a heliocentric velocity scale. The STIS data have an excellent velocity calibration with a zero point uncertainty of ~ 1 km s$^{-1}$ with occasional errors as large as 3 km s$^{-1}$ (see the appendix in Tripp et al. 2005). The S/N per 7 km s$^{-1}$ two pixel resolution element of the extracted spectrum for HE 0226-4110 is 11, 11, and 8 at 1250, 1500, and 1600 Å, respectively. The S/N in the STIS spectrum is substantially lower for $\lambda < 1180$ Å and $\lambda > 1650$ Å.

The ISM lines of S II, N I, O I, Fe II, Si II and Si III in the STIS spectrum have average observed heliocentric velocities of 11.3, 10, 10.7, 10.5, 9.3 and 8.2 km s$^{-1}$, respectively. These values are based on measurements of the following lines: S II $\lambda\lambda$ 1250.578, 1253.805, 1259.518; N I $\lambda\lambda$1199.500, 1200.223, 1200.710; O I $\lambda$1302.169; Fe II $\lambda$1608.451; Si II $\lambda\lambda$1304.370, 1526.707; and Si III $\lambda$1206.500. The ISM absorption lines that mostly trace neutral hydrogen regions (lines from S II, N I, O I, Fe II and Si II) have $<v_{HELIO}> = 10.4 \pm 1.4$ km s$^{-1}$. These observed velocities have been used to establish the zero point wavelength calibration of the FUSE observations (see §2.3).

A portion of the STIS spectrum extending from 1235 Å to 1255 Å is shown in Figure 1. The plotted spectrum (flux vs heliocentric wavelength) has been binned to the 7 km s$^{-1}$ STIS resolution. The spectral region illustrated contains the H I Ly β $\lambda$1025.722 and O VI $\lambda\lambda$1031.926, 1037.617 absorption lines from an intervening absorption system with $z_{abs}$ (O VI) = 0.20701. This particular system is the major focus of this paper. Various other absorption lines from the IGM and ISM are numbered on the figure and identified in the figure caption.

## 2.3. FUSE Observations

The FUSE observations of HE 0226-4110 were obtained over the time periods listed in Table 1. The 2000 and 2001 observations were obtained as part of the FUSE science team O VI project (Wakker et al. 2003). The 2002 and 2003 observations were obtained as part of the FUSE GO program DO27 to investigate the physical conditions in the highly ionized IGM. By combining these observations, a total exposure time of 193.3 ks was obtained. The QSO was aligned in the center of the LiF1 LWRS channel (30"x30"), which was used for guiding. The measurements were obtained in the time-tagged photon-address readout mode and cover the wavelengths from 916 to 1188 Å with a resolution of 20 to 25 km s$^{-1}$ (FWHM).

Discussions of the FUSE instrument and its inflight performance are found in Moos et al. (2000) and Sahnow et al. (2000). For each FUSE channel (LiF1, LiF2, SiC1, SiC2) a composite spectrum was produced incorporating all the valid observations for the channel. The FUSE spectra were processed with the FUSE pipeline software (CALFUSE v2.4.0). Our reduction steps mostly follow those previously described by Wakker et al. (2003) and Sembach et al. (2004). In brief, the numerous ISM lines seen in the spectrum of HE 0226-4110 were referenced to the velocities measured for the STIS ISM observations of S II, N I, O I, Fe II, and Si II in order to establish the velocity zero point offset for the different channel observations. The tracers of neutral hydrogen regions in the ISM toward HE 0226-4110 measured by STIS have average heliocentric absorption velocities of 10.4 ±1.4 km s$^{-1}$. The ionized ISM toward HE 0226-4110 absorbs at nearly the same velocity since v(Si III) = 8.6 km s$^{-1}$ (see §2.2). Forcing the ISM lines from the neutral and weakly ionized gas in the FUSE band to have v$_{HELIO}$ = 10.4 km s$^{-1}$ provides a good calibration for the zero point wavelength offsets.

A single velocity offset was applied to the individual observations in each detector segment based on measures of the ISM velocities in those observations. After all observations for a detector segment were co-added, we checked the final velocities of the ISM lines to see if small residual irregularities in the wavelengths existed. We found it necessary to shift the SiC2A segment observations near 930 Å by 4 km s$^{-1}$ to properly align the O I $\lambda\lambda$929.517, 930.256, and D I $\lambda$930.495 ISM absorption lines to the observed ISM neutral gas heliocentric velocity of 10 km s$^{-1}$. This region of the spectrum contains the important Ne VIII $\lambda\lambda$770.409 absorption line in the system at z =0.20701 (see §3).

The fully processed and calibrated spectra using the pipeline calibration (v2.4.0) with a careful zero point reference to the ISM absorption have a nominal velocity zero-point uncertainty of ∼ 5 km s$^{-1}$ (1σ). The relative velocity uncertainties across the full 916 to 1188 Å wavelength range are also ∼ 5 km s$^{-1}$ but may be larger near the edges of the detectors. The FUSE velocity calibration will be more reliable for those IGM absorption lines near ISM lines suitable for deriving the wavelength zero point offset.

The highly oversampled FUSE spectra were binned to a bin size of 4 pixels corresponding to ∼ 0.025Å or ∼ 7.5 km s$^{-1}$. This provides approximately three samples per 20 km s$^{-1}$ resolution element.

In the wavelength range from 916 to 987 Å the observations are from the FUSE SiC2A channel. Because of channel alignment problems, no observations were obtained in the SiC1B channel. Over the wavelength region from 987 to 1083 Å we used LiF1A channel observations. From 1087 to 1182 Å we used observations in the LiF2A channel. The lower S/N observations in the LiF1B channel were used to check for fixed pattern noise in the LiF2A observations. The final S/N achieved per resolution element near wavelengths of the important redshifted IGM lines in this study range from 8 to 17 to 17 near 950, 1000, and 1125Å, respectively.

Day and night observations were used except for the observations of O IV $\lambda$787.711 absorption for which only night data were used to reduce the effects of airglow O I contamination at 950.8Å. The FUSE spectrum of HE 0226-4110 is ideally suited for IGM studies because the contamination by the Galactic ISM $H_2$ absorption "forest" is relatively weak.

### 3. PROPERTIES OF THE z = 0.20701 ABSORPTION LINE SYSTEM

The FUSE and STIS high resolution observations of HE 0226-4110 shown in Figures 2a and 2b reveal the presence of a multi-phase absorption line system at $z_{abs}$(O VI) = 0.20701 containing absorption from H I (Ly $\alpha$ to Ly $\theta$), C III $\lambda$977.020, O III $\lambda$832.929, O IV $\lambda$787.711, O VI $\lambda\lambda$1031.926, 1037.617, N III $\lambda$989.799, Ne VIII $\lambda\lambda$770.409, 780.324, Si III $\lambda$1206.500, and S VI $\lambda$933.378. S V $\lambda$786.468 is tentatively detected. Important non-detections include C II $\lambda\lambda$903.962, 1334.532, N II $\lambda$1083.994, N V $\lambda$1242.804, O I $\lambda$1302.169, O II $\lambda$834.465, Si II $\lambda$1260.422, Si IV $\lambda$1402.773, S IV $\lambda$1062.664, and S VI $\lambda$944.523.

The line identifications were made as part of an effort to identify, measure, and analyze all the IGM and ISM absorption lines in the spectrum of HE 0226-4110. The full set of observations including all the identifications will be reported by Lehner et al. (in preparation). Fox et al. (in preparation) are analyzing the absorption produced by Milky Way high velocity gas in the direction of HE 0226-4110.

The continuum normalized absorption line profiles for the detections and important non-detections are displayed against restframe velocity in the z = 0.20701 reference frame of the O VI absorption in Figures 2a and 2b. Absorption lines with $\lambda_{rest}$ > 980 Å are from the STIS measurements. Those with $\lambda_{rest}$ < 980 Å are from the FUSE observations. In the case of H I Ly $\iota$ 920.963, N V $\lambda$1238.821, Si IV $\lambda$1393.760 and N IV $\lambda$765.147 our search for absorption by the ion is severely impacted by blending from ISM or IGM absorption from other ions. The O IV $\lambda$787.711 absorption is clearly recorded in the velocity range from –60 to 0 km s$^{-1}$ but blends with ISM O I $\lambda$950.885 for v > 0 km s$^{-1}$. Note that in the rest frame of the O IV $\lambda$787.711 observations shown in Figure 2b, the ISM O I absorption with a heliocentric velocity of 10 km s$^{-1}$ should appear at v = 45 km s$^{-1}$.

The continuum normalized velocity profiles were produced by fitting low order Legendre polynominals to each absorption feature. The basic measurements of absorption line velocity, line width, line strength, and column density are given in Tables 2 and 3. In Table 2 we list metal ion line measurements obtained by using the apparent optical depth method of Savage & Sembach (1991) where the continuum normalized flux as a function of velocity is converted into the apparent column density as a function of velocity, $N_a(v)$. The various quantities listed include the ion, the vacuum rest wavelength, the S/N in the continuum near the absorption line per FUSE or STIS resolution element, the absorption line average velocity,
$<v> = \int v\, N_a(v)dv/\, N_a$, the Doppler width of the absorption,
$b = [2 \int (v-<v>)^2\, N_a(v)dv/N_a]^{1/2}$, the line equivalent width in the rest frame, $W_{rest}$, and the log of the apparent column density, $logN_a$, where $N_a = \int N_a(v)dv$. The integration ranges for calculating these quantities for the metal lines are shown with the horizontal bar above the absorption profiles displayed in Figures 2a and 2b. We also list E140M for observations obtained with STIS and the segment for observations obtained with FUSE. Notes to Table 2 provide information about blending problems or other difficulties associated with the measurements.

The 1σ errors on the observed quantities listed in Table 2 include the effects of the random noise and the continuum fit error added in quadrature (see Wakker et al. 2003). The method of fitting Legendre polynominals produces an error in the coefficients that is used to estimate the continuum placement error (see Sembach & Savage 1992). Fixed pattern noise errors for the FUSE measurements are expected to be small for this data set because of the pattern noise averaging that occurs when a very large number of individual spectra obtained at somewhat different detector positions are combined to produce the final co-added spectrum. The FUSE and STIS background levels are very reliably determined for objects as bright as HE 0226-4110 except at wavelengths contaminated by terrestrial airglow features. For the non-detections we list the 3σ upper limit to the restframe equivalent width and derive the 3σ upper limit to the column density by assuming the absorption lines lie on the linear part of the curve of growth. Rest frame wavelengths and f-values used in our study are from Morton (2003) for $\lambda > 912$Å and from Verner et al. (1996) for $\lambda < 912$ Å.

The O VI $\lambda\lambda$1031.926, 1037.617 lines are detected with high significance with $W_{rest} = 169\pm15$ and $112\pm10$ mÅ. This implies a detection significance of 11.3σ and 11.2σ for the two members of the O VI doublet and a combined doublet detection significance of 15.9σ assuming the errors are dominated by photon counting statistics. The S VI $\lambda$933.378 absorption line is in a blend free region of the spectrum. We measure $W_{rest} = 27.0\pm8.2$ mÅ which is a 3.3σ detection. However, the weaker $\lambda$933.378 member of the S VI doublet is not detected with $W_{rest} <16.6(3\sigma)$ mÅ. The S V $\lambda$786.468 line is tentatively detected with $W_{rest} = 22.7\pm10.2$ mÅ corresponding to a significance of 2.2σ. A 3σ upper limit for S V is $W_{rest} <30.6$ mÅ(3σ).

### 3.1. Absorption by Ne VIII at z = 0.20701

The first detection of the intergalactic Ne VIII doublet is shown in Figure 2b and in Figure 3 where the measurements are shown over an extended velocity range to better illustrate the continuum placement and the noise levels of the spectra. Four pixel binning (= 7.5 km s$^{-1}$) was used for the Ne VIII measurements in both figures. In the upper panel of Figure 3 the day+night measurements are shown as the solid histogram while the night-only measurements are shown as the red histogram. The night-only integration time is ~ 50% of the total integration time.

The wavelength region containing the weaker Ne VIII $\lambda$780.324 line at z = 0.20701 is shifted to 941.859Å in the FUSE spectrum. This wavelength region does not contain close-by interfering absorption or emission lines from the ISM or the terrestrial atmosphere except for a very weak possible contribution from the H$_2$ Lyman (15-0) P(2) $\lambda$941.596 line discussed below .

The stronger Ne VIII $\lambda$770.409 line is redshifted to 929.891Å in the FUSE spectrum. This wavelength region is affected by ISM absorption from O I

λλ929.517, 930.256, D I λ930.495, H$_2$ Werner band (4-0) R(0) absorption at 929.532Å, H$_2$ Werner (4-0) R(1) absorption at 929.688Å, and possibly by H$_2$ Lyman band (17-0) P(3) absorption at 929.687Å.

The ISM O I lines are expected to be strong. In the reference frame of the Ne VIII λ770.409 panel, the O I absorption lines at v$_{HELIO}$ = 10 km s$^{-1}$ are expected to occur at -113 and 126 km s$^{-1}$, respectively. These two O I absorption lines are identified in Figures 2b and 3. A comparison of the black (day+night) and red(night-only) observations in the upper panel of Figure 3 reveals that the continuum to each side of the O I λ929.517 absorption in the day+night observations is affected by terrestrial O I airglow emission while O I airglow emission is not apparent for the O I λ930.256 line in the day+ night observations. H I airglow emission substantially fills in the ISM H I λ930.748 absorption shown in Figure 3 for the longer integration day+night measurements.

Fortunately, the ISM O I absorption lines are outside the –80 to +80 km s$^{-1}$ velocity range where we might expect to see the Ne VIII λ770.409 IGM absorption. The ISM D I λ930.495 absorption is expected to occur at 203 km s$^{-1}$ and is marked on the panels in Figures 2b and 3. The value of D/H in the low velocity ISM toward HE 0226-4110 will be the topic of a separate investigation.

In order to evaluate possible H$_2$ contamination near the Ne VIII λ770.409 absorption we have identified and measured equivalent widths for all the H$_2$ ISM lines in the spectrum of HE 0226-4110. The identifications will be included as part of the publication reporting the complete analysis of the HE 0226-4110 IGM spectrum (Lehner et al. in preparation). The H$_2$ absorption toward HE 0226-4110 is relatively weak with detected lines arising from H$_2$ in the rotation levels J = 0, 1, 2, and 3.

Absorption by ISM H$_2$ occurs at a heliocentric velocity of 10 km s$^{-1}$. Therefore, in the rest frame of the Ne VIII λ770.409 line, the H$_2$ Werner (4-0) R(0) λ929.532, H$_2$ Werner (4-0) R(1) λ929.688, and H$_2$ Lyman (17-0) P(3) λ929.687 absorptions lines are expected to occur at –108, -57, and –58 km s$^{-1}$, respectively.

A curve of growth analysis of the reasonably well detected H$_2$ transitions out of the J = 0, 1, 2 and 3 levels allows us to estimate values of logN(H$_2$, J). We obtain b = 6 km s$^{-1}$ and logN(H$_2$, 0) = 13.85; logN(H$_2$, 1) = 14.06; logN(H$_2$, 2) = 13.44; and logN(H$_2$,3) = 13.51, with errors of ~±0.3 dex on the column densities of the J=0 and 1 levels and ~+0.5, -1.2 dex on the J = 2 and 3 column densities. The error on b is poorly constrained. These results allow us to predict the expected equivalent widths of the three H$_2$ lines near the Ne VIII λ770.409 line. In the rest frame of the Ne VIII absorption, the predicted equivalent widths for the H$_2$ (4-0) R(0) λ929.532, (4-0) R(1) λ929.688, and (17-0) P(3) λ929.687 lines are 15.2 mÅ, 11.9 mÅ, and 0.6 mÅ, respectively. The errors on these equivalent widths are ~ +100% and –50%. The absorption profiles of these lines convolved with the FUSE 20 km s$^{-1}$ line spread function are displayed on the continuum in Figure 3. The (4-0) R(0) λ929.532 absorption occurs at –108 km s$^{-1}$ and severely blends with ISM O I 929.517 absorption. The (4-0) R(1) λ929.688 and (17-0) P(3) λ929.687 lines with a total equivalent width of 12.5 mÅ at –57 km s$^{-1}$ lie on the negative velocity side of the Ne VIII λ770.409 absorption but don't significantly affect our ability to study the absorption. Various other much weaker H$_2$ absorption lines are also shown on the two panels of Figure 3. The only other line of possible importance to the Ne VIII study is the Lyman (15-0) P(2) $\lambda941.596$ line which lies at $-76$ km s$^{-1}$ in the Ne VIII $\lambda780.324$ reference frame. However, that $H_2$ line is expected to have a very small equivalent width of only $W_\lambda = 0.7$ mÅ.

Based on these predicted $H_2$ line strengths we conclude that the $H_2$ ISM absorption is not interfering with our ability to study the Ne VIII $\lambda770.409$ line in the absorption system at $z = 0.20701$ and the Ne VIII $\lambda780.324$ line is nearly blend free.

For the two Ne VIII $\lambda\lambda770.409$, 780.324 lines we measure rest frame equivalent widths of $32.9\pm10.5$ and $24.9\pm10.6$ mÅ over an integration range from $-50$ to 42 km s$^{-1}$. The integration range is only slightly smaller than the range $-55$ to $+55$ km s$^{-1}$ adopted for the stronger O VI lines. The measured equivalent widths and errors correspond to a formal detection significance of $3.1\sigma$ and $2.3\sigma$, respectively, for each Ne VIII line. Note that the detection significance for each Ne VIII line increases to $3.2\sigma$ and $2.9\sigma$ if the velocity range of the equivalent width integration is narrowed to $-52$ to $+20$ km s$^{-1}$. The $2.3\sigma$ feature near v = 0 km s$^{-1}$ that we attribute to Ne VIII $\lambda780.324$ in the lower panel is the strongest feature in the velocity range from $-450$ to 700 km s$^{-1}$. For comparison, the weaker unidentified feature at $+380$ km s$^{-1}$ has a lower detection significance of $1.5\sigma$ when integrated over a 90 km s$^{-1}$ wide velocity range.

The relative strength of the lines in the Ne VIII panels near v = 0 km s$^{-1}$ in Figures 2b, 3, and 4 is consistent with the Ne VIII identification. It is difficult to find any reasonable alternative identifications from the ISM or from metal or H I lines associated with IGM absorbers at other redshifts in the spectrum of HE 0226-4110. The good velocity alignment of the two Ne VIII absorption lines with respect to each other and with respect to the O VI absorption gives us added confidence that the Ne VIII has been detected in the $z = 0.20701$ system.

Since both lines have been detected, the detection significance of the Ne VIII doublet is greater than the individual line detection significance given above. If the errors are dominated by photon count statistics, the $3.1\sigma$ and $2.3\sigma$ individual line significance implies a joint detection significance for the doublet of $3.9\sigma$.

Another way to determine the joint detection significance is through the simultaneous Voigt profile fit to the Ne VIII doublet discussed in §3.3. The simultaneous profile fit result of logN(Ne VIII) = $13.89\pm0.11$ corresponds to a $4.5\sigma$ detection significance for Ne VIII at $-7\pm6$ km s$^{-1}$. A similar simultaneous fit to the O VI doublet yields logN(O VI) = $14.37\pm0.03$ corresponding to a joint $15\sigma$ detection significance for the two components of the O VI doublet at $0\pm2$ km s$^{-1}$. Ne VIII and O VI absorb at the same velocity to with in the velocity error.

The different reduction methods yield somewhat different values for the significance of the detections. The simultaneous profile fit process imposes the assumption that simple one-component Voigt profiles describe the two Ne VIII lines. This additional constraint on the absorption may contribute to the larger significance determined for the profile fit result ($3.9\sigma$ versus $4.5\sigma$). Other reasons could include differences in the measuring software with respect to the continuum placement and the continuum error estimates.

An alternate way to determine the joint detection significance for both lines in the Ne VIII doublet that does not require an assumption about line shape is to co-add in velocity space the measurements for the weak line and the strong line shown in the two panels of Figure 3. In the co-added spectrum the rest equivalent width for the Ne VIII line is found to be 31.5±6.9 mÅ over the –50 to +42 km s$^{-1}$ velocity range when using the software that allows for statistical errors and the continuum placement error. The co-added absorption line has a detection significance of 4.6σ. Assuming this co-added weak line is on the linear part of the curve of growth, this equivalent width corresponds to logN(Ne VIII) = 13.89±0.09 when assuming the effective f-value for the co-added absorption is the average of the f-values for the Ne VIII λλ770.409, 780.324 lines. This column density and error are very close to the profile fit result of 13.89±0.11.

The level of significance determined for the Ne VIII absorption doublet depends on the measuring and error estimating techniques we have employed. However, in all three cases the doublet detection significance is found to lie between 3.9σ and 4.6σ, We conclude with confidence that the Ne VIII absorption doublet is detected in the z = 0.20701 system at a velocity that is essentially the same as found for O VI.

### 3.2. Component Fits to the Lyman Series Absorption

The H I Lyman series lines shown in Figure 2a reveal H I is detected in absorption from Ly α λ1215.670 to Ly κ λ923.150. The stronger lines of Ly α λ1215.670 to Ly δ λ949.743 reveal that the absorber has more than one component. The principal component lies near –25 km s$^{-1}$ with a weaker component near 5 km s$^{-1}$. To separate the absorption in these two H I components we fitted Voigt profiles to the absorption. The best fit result is given at the top of Table 3 and is based on a simultaneous two component profile fit to the Ly α, Ly β, and Ly γ lines from STIS and to the Ly γ to Ly κ lines from the FUSE LiF2A channel observations. The fit also includes the FUSE LiF1B channel H I observations for Ly γ, Ly δ, and Ly ε. The fit was performed using the Voigt component fitting software of Fitzpatrick and Spitzer (1997) and assumes the FUSE instrumental spread function is a Gaussian with a FWHM = 20 km s$^{-1}$. For the STIS instrumental spread function we adopted the two component profile from the STIS Instrument Handbook (Proffitt et al. 2002). Fitting the STIS observations of Ly α, β, and γ independently produces the results listed as the second set of results for H I in Table 3. The results from a full fit of the FUSE and STIS observations are quite similar to the results from a fit to the STIS data alone, although the inclusion of higher Lyman lines better constrains the column density of the strong component. This illustrates the value of having FUSE observations of the higher Lyman lines for determining accurate values of N(H I).

The best fit result for H I listed at the top of Table 3 using the STIS and FUSE spectra has a reduced chi square, $\chi_\nu^2$ = 0.906 and reveals the presence of two absorbing components with v = -24±1 and 5 ±2 km s$^{-1}$; b = 17.4±1.4 and 35.9±1.1 km s$^{-1}$; and logN(H I) = 15.06±0.04 and 14.89±0.05. The column densities of the two components differ by a factor of 1.48 while the b values differ by a factor of 2.06. The best fit profiles are compared to the H I absorption line observations in Figure 4 where the vertical tick marks above the profiles indicate the velocities of the two components.

The H I fit parameters are not as accurately determined as the formal errors coming from the Voigt component fit process might suggest. The validity of the fit results and their errors depend critically on the validity of the simplistic assumption that both absorbers are well described by single Voigt optical depth profiles. If the true nature of the absorption was somewhat more kinematically complex, the derived numbers could change considerably. The low S/N STIS observation of the H I Ly $\gamma$ $\lambda$972.537 line (not shown in Figs 2 or 4) suggests a possible additional narrow H I component near 17 km s$^{-1}$. However, the measurements are too noisy to establish if the suggested component is real.

The fit results listed in Table 4 were obtained to evaluate the sensitivity of the H I component fits to input assumptions involving the velocity of the principal component. For the six different fit results listed, we fixed the velocity of the principal absorption component and allowed the five other fit parameters to vary until a best fit was obtained. The resulting reduced chi square, $\chi_v^2$ is listed for each of the six cases along with the best fit results for v, b, and logN for each component. The overall best fit is obtained for a velocity of the principal component of –24 as revealed by the fully free fit listed in Table 3. However, note that quite good values of $\chi_v^2$ are also obtained for all the other cases fitted. The fit results for principal component velocities of –27 and –20 km s$^{-1}$ are hardly distinguishable from the best fit case of –24 km s$^{-1}$. In Figure 4 we have added to the fitted profile for Ly $\delta$ with the dotted and dashed lines the fit cases where the velocity of the principal component is –30 and –17 km s$^{-1}$, respectively. It is only for velocity differences this large that one can see with the eye the beginning of the deterioration in the quality of the fit. In the following discussions we will adopt the best fit results for H I listed at the top of Table 3 but will attach systematic errors to the H I fit results to better reflect the difficulty of assigning accurate parameters to overlapping absorbers with roughly similar strengths and widths. For the stronger component #1 we adopt the parameters v = -24±1 ±4 km s$^{-1}$, b = 17.4±1.4(+3.6, -2.3) km s$^{-1}$, and logN(H I) = 15.06±0.04(+0.10, -0.05), where the second error is our estimate of the ±1σ systematic error associated with the fit process. For the weaker component #2 we adopt v = 5 ±2 (+8, -2) km s$^{-1}$, b = 35.9±1.1(+0, -2.2) km s$^{-1}$, and logN(H I) = 14.89±0.05 (+0.11, -0.20). The systematic errors on the second weaker component are large and asymmetric.

In §5 as part of the discussion of the origin(s) of O VI and Ne VIII in hot collisionally ionized gas we explore the possibility that there may exist a very broad Ly $\alpha$ absorption centered at the velocity of the O VI absorption. Broad Ly $\alpha$ absorption lines with b values ranging from 50 km s$^{-1}$ to more than 100 km s$^{-1}$ have been previously seen in high quality STIS observations of three QSOs (Tripp et al. 2001; Richter et al. 2004; Sembach et al. 2004). The broad Ly $\alpha$ absorption is easier to recognize when it is not associated with multiphase absorbers.

### 3.3. Component Fits to the Metal Line Absorption

Voigt profiles were also fitted to the metal lines. The results are listed in Table 3 and shown in Figure 4. Single Voigt optical depth components were fitted to the single absorption lines of Si III $\lambda$1206.500, and N III $\lambda$989.799. For O VI $\lambda\lambda$1031.926, 1037.717 , Ne VIII $\lambda\lambda$770.409, 780.324, and S VI $\lambda\lambda$933.378, 944.523,

the two components of the doublets were simultaneously fitted with single Voigt optical depth components. The inclusion of the non-detected S VI λ944.523 line in the fit process explains why the S VI column density derived from the profile fit is 0.15 dex smaller than the value listed from the direct $N_a(v)$ integration in Table 2.

The very strong line of C III λ977.020 shows an asymmetric extension to positive velocity implying the weaker component seen in H I may also exist  in C III.  The O III λ832.929  line also reveals a hint of this second component.  For C III and O III, the Voigt profile fit was performed assuming two components fixed in velocity at –15 and 10  km s$^{-1}$.  The –15  km s$^{-1}$ velocity for the principal absorber was adopted from the STIS results for Si III λ1206.500 where the S/N is relatively high and the STIS velocity calibration is excellent.  The velocity of 10  km s$^{-1}$ for the second component was obtained from a two component free fit to the O III measurement with the velocity of the main component fixed at –15  km s$^{-1}$.  The C III absorption is so strong that a two component free fit placed the second component velocity close to the main component which is incompatible with the O III observation and the result for the weaker metal line of Si III.  The component fitting procedure we adopted for C III and O III was designed to best determine the amount of  C III and O III absorbing at velocities where Si III and N III are detected.  Although this procedure is not ideal, it hardly affects the final result.   The C III absorption is so strong that the dominant error is associated with line saturation effects rather than the component separation problem.

## 4.  THE MODERATELY IONIZED GAS ABSORPTION

We associate the absorption by C III, N III, O III, and Si III near –15 km s$^{-1}$ with the primary H I absorption component with logN(H I) = 15.06±0.05(+0.10,-0.05), b = 17.4±1.4(+3.6,-2.3) km s$^{-1}$ and  v = –24 ±1±4  km s$^{-1}$.  The most reliable metal line velocity for this absorber is that provided by the STIS Si III observation of v = –15 ±3 km s$^{-1}$.  The observations of the other metal lines including C III, N III, and O III are consistent with the principal moderate ionization absorption being near v = -15 km s$^{-1}$ (see Table 3  and Fig. 4).   Although the formal best fit velocity for the principal H I component is –24±1 km s$^{-1}$, we have noted in §3 that the  $\chi_v^2$ of the two component fit to the H I observations has a broad minimum such that the fits for the principal component at –20, –24 and –27 km  s$^{-1}$ are hardly distinguishable.  We therefore associate the moderate ionization absorption with the H I in the principal absorbing component and explore the implications of that association.

We adopt the following column densities  and errors  for the moderate ionization absorber taken from Table 3:  logN(H I) = 15.06±0.04 (+0.10, -0.05),  log N(C III) = 14.04±0.22:, logN(N III) = 13.61±0.16,  logN(O III) = 14.23±0.07,  and logN(Si III) = 12.46±0.08.  Note that the true error for C III is probably larger than that quoted because the line is so strongly saturated (see Figs. 2 and 4).  Upper limits for C II,  N II, O I, O II,  Si II, and Si IV are taken from Table 2.

As part of our photoionization modeling we consider the possibility that some or all of the detected O IV and S VI and the tentatively detected S V may be associated with the moderate ionization absorber.   We adopt logN(O IV) >14.50  from Table 2 based on the    $N_a(v)$ integration of the blended O IV line from –60 to 0 km s$^{-1}$ and

logN(S VI) = 12.78±0.11 from Table 3 based on the profile fit result.  For S V we take logN(S V) = 12.49 (+0.16, -0.28) from Table 2 based on the tentative 2.2σ detection of this ion.  We note in §5 that a photoionization origin for the bulk of the O VI and Ne VIII is very unlikely.

We consider whether the absorption by the moderate ionization species could be produced in a low-density photoionized medium.  We therefore calculated photoionization models for a  plane parallel distribution of low density gas with the ionization code CLOUDY (V94.00;  Ferland et al. 1998) in order to compute column densities of various ions through a slab illuminated with the Haardt & Madau (1996) UV background radiation field from QSOs and AGNs for a redshift of 0.2.  At 912Å the mean specific intensity in the assumed radiation field is  $J_v = 1.9 \times 10^{-23}$ erg cm$^{-2}$ s$^{-1}$ Hz$^{-1}$ster$^{-1}$.  The model assumes solar relative heavy element ratios from Grevesse & Sauval (1998) using the recent updates from Holweger (2001) for N and Si,  from Allende Prieto, Lambert, & Asplund (2002) for C, and from Asplund et al. (2004) for O and Ne.  The actual adopted abundances are listed in Table 5 where the updated values  are compared to the values given recently by Grevesse & Sauval (1998).  Note that the solar reference abundances of C, N, O,  and Ne have all undergone considerable changes over the last several years.  It is therefore important to make sure that the same reference abundances are used when comparing observations to theoretical predictions published in different papers.

The best fit model is shown in Figure 5 where predicted column densities are plotted against the log of the ionization parameter,   U  = n$_\gamma$/ n$_H$, where n$_\gamma$ is the hydrogen ionizing photon density and n$_H$ is the total hydrogen (H°+ H$^+$) density.  The model shown assumes logN(H I) = 15.06 and best fit is achieved with [Z/H]= -0.50.  In this paper we adopt the standard abundance notation where [Z/H] = log (Z/H) –log(Z/H)$_O$ where Z refers to  a heavy element. The different symbols show the expected run of logN(ion) vs log U for many species.   For all the detected species the heavy black lines denote the range of logU over which the model predictions are in agreement with the observations including the formal  ±1σ errors or the lower limit for O IV.   Large values of  logU are required to explain the measurements of O VI and Ne VIII while the observations for O IV and S VI  can either be explained by large or moderate values of logU.

The column densities of C III, N III, O III, Si III, O IV, and S VI are in agreement with this model for log U = -1.85 with the strongest constraints being provided by the measurements for O III , Si III,  O IV, and S VI.  The model is also consistent with the column density for S V based on the tentative detection of a single line.  When we first explored this fit we did not expect the model fit to also explain most of the observed S VI and O IV.    However, as we will see in §5 it is unlikely that these species are associated with the hot gas detected in the the lines of O VI and Ne VIII.

The photoionization model fit result shown in Figure 5 implies  that the absorber has a solar abundance ratio for C, N, O,  Si, and  S and an overall metallicity [Z/H]  = -0.50.  This particular model fit also satisfies the column density upper limits for C II, N II, O I, O II, Si II, and Si IV listed in Table 2.

We note that the   model with logU = -1.85 predicts logN(O VI) = 13.15 and a negligible amount of Ne VIII.  The total observed amount of O VI, which is mostly produced in hot gas (see §5), exceeds that expected from photoionization for log U =-1.85 by a factor of 17.

The properties of the model absorber required to fit the observations of H I, C III, N III, O III, Si III, O IV, and S VI, and the limits on C II, O I, O II, N II, Si II, and Si IV are listed in Table 6. The temperature of $2.1 \times 10^4$ K derived for the gas is consistent with the temperature limits derived from the breadth of the H I absorption. By combining the $1\sigma$ statistical error and the systematic error, the $1\sigma$ upper limit to b(H I) is 22.4 km s$^{-1}$ which implies an upper limit to T of $<3.0 \times 10^4$ K. The gas has a low total hydrogen density, $n_H = 2.6 \times 10^{-5}$ cm$^{-3}$, and low pressure, P/k = 0.54 cm$^{-3}$ K. The gas is highly ionized with N(H I)/N(H) = $2.5 \times 10^{-4}$. A path length of 57 kpc is required to produce N(H) = $4.6 \times 10^{18}$ cm$^{-2}$.

It is difficult to assign reliable errors to the model fit results listed in Table 6 because the fit results are not unique. The observations are well matched with abundances in the absorbing gas with solar ratios of C, N, O, Si, and S with the strongest constraints provided by Si III and S VI. Deviations from the solar ratios among the heavy elements would result in shifts of the best fit value of logU and currently there is a 0.13 dex discrepancy between the photospheric and meteoric value of the Solar S abundance (see Table 5) The H I reference column density used for the fit may have random plus systematic errors as large as 0.15 dex (see Table 4). The shape and absolute level of the assumed extragalactic ionizing background we have adopted from Haardt & Madau (1996) are uncertain (see Collins et al. 2003 for a discussion of the problem). Changing the hydrogen ionizing photon density, $n_\gamma$, by a factor of several will modify the estimated value of $n_H$ by a similar factor since the ionic ratios are determined by $U = n_\gamma / n_H$. A modification of $n_H$ will in turn introduce modifications to P/k and L. A modification to the shape of the spectrum will introduce shifts into the ion curves shown in Figure 5 and would lead to changes in the elemental abundances required to match the observations to the theoretical curves. Exploring all these photoionization modeling uncertainties is well beyond the scope of this paper.

Given these considerations and plausible uncertainties associated with the photoionization model fitting process, it appears that a fair statement would be that the observations are consistent with an overall metallicity of [Z/H] = $-0.50 \pm 0.20$ dex and solar abundance ratios to within 0.15 dex, except for C where the uncertainty is larger because of absorption line saturation effects. The uncertainties in U, N(H), and P/k are also $\sim \pm 0.15$ dex. However, the errors on T and L are $\sim \pm 0.05$ dex and $\sim \pm 0.3$ dex, respectively.

The derived values of metallicity, density, pressure, and path length are suggestive of an origin of this photoionized absorber in the gas of a galaxy group or in the very distant outskirts of a galaxy halo. Unfortunately, at the current time there are very few galaxy redshift measurements for the galaxies in the general direction of HE0226-4110 so it is not possible to associate the z = 0.20701 absorber with the environment of a particular galaxy or group of galaxies. This problem will be corrected in a few years with several on going programs to determine the redshifts of galaxies in the direction of those QSOs for which high quality STIS and FUSE absorption line measurements have been obtained.

Several low redshift photoionized metal line systems have been previously studied. For the systems at z = 0.06807 toward PG 0953+414 (Savage et al. 2002), z = 0.29236 toward PG 1259+593 (Richter et al. 2004), and for several absorbers toward PKS 0405-125 (Prochaska et al. 2004) it is interesting that the observed metallicities and required values of logU, $n_H$, and L are roughly similar to the values listed in Table 6. In the case of the system at z = 0.06807 toward PG 0953+414, the galaxy redshift survey of Savage et al. (2002) reveals that the absorber does occur near a strong enhancement in the density distribution of galaxies as a function of redshift along the general line of sight. The multiphase aspect of the z = 0.20701 absorber has also been found for other low redshift metal line absorbers including the z = 0.05104 and z = 0.06438 absorbers toward PG1211+143 ($z_{em}$ = 0.081) studied by Tumlinson et al. (2005)

## 5. THE HIGHLY IONIZED GAS ABSORPTION

The production of O IV, S VI, O VI, and Ne VIII from their next lower ionization states requires photons or electrons with energies of 54.9, 72.7, 113.9, and 207.3 eV, respectively. In CIE, O IV, S VI, O VI and Ne VIII peak in abundance at T = $1.6 \times 10^5$, $1.8 \times 10^5$, $3.0 \times 10^5$, and $7.0 \times 10^5$ K, respectively. Therefore, the detection of these ions in the z = 0.20701 absorption system toward HE 0226-4110 with column densities of logN(O IV) >14.50, logN(S VI) = 12.78±0.11, logN(O VI) = 14.37±0.03, and logN(Ne VIII) = 13.89±0.11 (see Table 3) is suggestive of the presence of hot collisionally ionized gas. However, photoionization in a very low density medium is an alternate possibility and we saw in §4 that most of the O IV and S VI is likely produced by photoionization in the absorber containing C III, N III, O III, and Si III.

The O VI doublet absorption profile is well fitted ($\chi_v^2$ = 1.04) with a Gaussian optical depth absorbing component with b(O VI) = 31.4±2.0 km s$^{-1}$ which constrains the temperature of the O VI absorbing gas to be T < $9.5 \times 10^5$ K. The profile breadths of the S VI and Ne VIII absorption lines are too uncertain to provide additional meaningful temperature constraints for the temperature of the gas producing these ions.

### 5.1. Production of the Highly Ionized Atoms by Photoionization

Before directly beginning to interpret the O VI and Ne VIII observations with models of gas at high temperature it is of interest to show that these absorption lines are unlikely to arise in very low density photoionized gas. We therefore created the photoionization model shown in Figure 6 (similar to the model discussed in §4) to evaluate the parameters required to produce O VI and Ne VIII in the photoionized gas. We again adopted the Haardt & Madau (1996) QSO and AGN extragalactic radiation field for z = 0.2 and calculated the expected behavior of O IV, O VI, S VI and Ne VIII as a function of logU. In the model we set [Z/H] = -0.5 to be the same as that found for the moderate ionization absorber (see §4) and then chose logN(H I) = 13.67 in order to fit the values of logN(O VI) and logN(Ne VIII). Note that increasing or decreasing logN(H I) can be compensated by decreasing or increasing [Z/H] in the model fit.

One can see with reference to Figure 6 that the observed column densities of O VI and Ne VIII, logN(O VI) = 14.37±0.03 and logN(Ne VIII) = 13.89 ±0.11, require log U = -0.09 and the small density $n_H \sim 4.5 \times 10^{-7}$ cm$^{-3}$. The model with logU = -0.09 predicts column densities of O IV and S VI that are ~2 and ~3 dex smaller than the observed column densities. This is not a problem because O IV and S VI appear to arise in the moderately ionized gas producing C III, N III, O III, and Si III (see §4).

The photoionization model using the expected extragalactic background radiation field is not likely the correct explanation for the observed O VI and Ne VIII because at the large required value of logU, the gas density is so small that the required path length is ~11 Mpc. Over such a large path, the Hubble flow broadening of the absorption would create a line with a breadth of ~ 760 km s$^{-1}$ (see the bottom scale on Fig. 6) which is ~10 times wider than the observed high ionization line absorption. If the absorbing structure producing the O VI and Ne VIII is illuminated by the relatively weak general extragalactic background radiation field at z = 0.2, the gas must be hot and ionized by electron collisions.

We have assumed in this discussion that the absorber is an intervening system and does not arise in the fast outflow from the QSO itself. If the absorption did arise in an outflow, the intensity of the ionizing radiation would be vastly larger than for the general ISM and the Hubble flow comments given above would not be valid. Although we can't exclude the possibility that the absorber is an outflowing associated absorption system, it is noteworthy that the observed absorption lines are relatively narrow (< 100 km s$^{-1}$) for an outflow where the implied ejection velocity is $5.3 \times 10^4$ km s$^{-1}$. Moreover, the absorber at z = 0.20701 does not show the clear signatures of ejected gas that are evident in low-redshift examples of QSO outflows (e.g. Yuan et al. 2002; Ganguly et al. 2003) such as partial covering of the flux source. The strong lines are black in their cores which indicates that the absorber completely covers the flux source as expected for an intervening system. A search for galaxies possibly associated with the absorber and/or measures of variability in the absorber would help to discriminate between the intervening and outflowing associated system origins for the system at z = 0.20701.

*5.2. Production of the Highly Ionized Atoms by Collisional Ionization in Hot Gas*

Since a photoionization origin for the Ne VIII and O VI in the general IGM has been ruled out (see §5.1), we turn to ionization mechanisms involving collisional ionization in a hot gas. The simplest hot gas model we can consider for explaining the O VI and Ne VIII absorption is for gas in CIE. In Figure 7 we show a CIE model displaying predicted column densities for H I, O IV, O VI, S VI, and Ne VIII for a uniform slab of gas with logN(H) = 19.92, [Z/H] = -0.50 and Solar abundance ratios for C, O, Ne, and S. The CIE model results for the ionic fractions have been taken from Sutherland & Dopita (1993). We have adopted the same metallicity as found for the photoionized absorber discussed in §4. The observed column densities of O VI (logN= 14.37±0.03) and Ne VIII ( logN=13.89±0.11) in the gas are plotted on the model curves at a value of log T = 5.73 where the observed and predicted values of N(O VI) and N(Ne VIII) agree. In the discussion above we have not reduced logN(O VI) to allow for a logN(O VI) = 13.15 contribution from the photoionized gas modeled in §4. With that correction logN(O VI) in the hot gas is 14.34 dex rather than 14.37 dex and the derived temperature hardly changes.

At T = $5.4 \times 10^5$ K the predicted column densities for O IV and S VI are significantly smaller than the observed column densities. This is consistent with the O IV and S VI arising in the photoionized gas phase containing C III, N III, O III, and Si III.

The CIE model shown in Figure 7 with logN(H)= 19.92 and [Z/H] = -0.50 has logN(H I) = 13.69 for logT=5.73. The neutral hydrogen associated with this hot gas will have a substantial amount of thermal line broadening. The expected Doppler parameter for H I is 95 km s$^{-1}$. A Ly$\alpha$ line with this value of N(H I) and b is in principle detectable. However, it would be superimposed on the narrower H I Ly$\alpha$ absorbers in the z = 0.20701 multiphase absorption line system. The broad H I absorber would be expected to have v = v(O VI) ~ 0 km s$^{-1}$, logN(H I) ~ 13.69, and b ~ 95 km s$^{-1}$. We have studied the STIS H I Ly$\alpha$ observations looking for evidence of a weak but broad H I component superposed on the narrower absorbing components discussed in §3. Figure 8 shows the results of that study. The histogram in Figure 8a shows the STIS Ly$\alpha$ measurements and the solid line shows the two component fit to the STIS Ly$\alpha$, Ly $\beta$, and Ly $\gamma$ observations as reported in Table 3 (second entry for H I). Figure 8b shows the results of a free fit obtained by adding a third broad component into the profile fitting process. The best fit result shown in Figure 8b has two narrow components with parameters nearly identical to those given in Table 3 and a third broad component with logN(H I) = 13.31±0.39, b =97: km s$^{-1}$ and v= -7: km s$^{-1}$. The errors on this broad component are large and the detection significance is low (~1.7$\sigma$). However, the parameters for a possible broad H I component derived from this free fit to the data are quite similar to the values expected from the O VI and Ne VIII measurements. Figure 8c shows a broad component with b = 97 km s$^{-1}$, v = -7 km s$^{-1}$, and logN(H I) = 13.70 which is 1$\sigma$ larger than for the best fit value of logN(H I) for the broad component in Figure 8b. Although we do not claim the detection of a broad component, we note that a broad component with logN(H I) in the range from 13.3 to 13.7 is compatible with the STIS observations.

The expected value of logN(H I) in the broad component depends on the assumed metallicity of the hot gas and the value of logN(H). Increasing the value of [Z/H] for the CIE model in Figure 7 from –0.50 to –0.20 changes the prediction for the expected column density in a broad component from logN(H I) = 13.69 to 13.39 and decreases logN(H) to 19.72. The observations of O VI, Ne VIII, and the possible presence of a broad component of H I are therefore consistent with CIE conditions in a hot gas with [Z/H] > -0.5, solar relative heavy element abundances of O and Ne, log T= 5.73, logN(HI) < 13.7 and logN(H) < 19.92.

A hot gas model for the origin of the Ne VIII and O VI absorption can be constructed so that it does not violate constraints imposed by Hubble flow broadening. For the hot gas model, the level of ionization is not controlled by the gas density as in the case of the photoionized gas model discussed and rejected in §5.1. The hot gas model is consistent with broadening from the Hubble flow so long as the path length through the absorber is less than ~1 Mpc which would broaden the Ne VIII and O VI lines to ~70 km s$^{-1}$. This 1 Mpc depth constraint coupled with logN(H) = 19.92 for CIE and [Z/H] = -0.5 implies the total hydrogen physical density in the absorber is n(H) $\geq 2.7 \times 10^{-5}$ cm$^{-3}$. If the absorber has a similar extent in all three dimensions its volume is ~1 Mpc$^3$ and the total mass is ~9x10$^{11}$ M$_O$ for He/H = 0.085 and n(H) = 2.7x10$^{-5}$ cm$^{-3}$. However, this mass estimate is very uncertain since we have no information about the true shape of the absorber.

Gas in the temperature range from 10$^5$ to 6x10$^5$ K can cool relatively rapidly if its density is large since this is the temperature range where the cooling rate of an ionized plasma is a maximum. In fact O VI and Ne VIII are the major coolants in this temperature range. It is therefore reasonable to question the general validity of CIE models for the ionization of the hot gas.

It is unlikely the O VI and Ne VIII absorption is due to the non-equilibrium ionization occurring in conductive interfaces between cool gas and a hot unseen exterior gas. A typical conductive interface will produce a column density of O VI of ~10$^{13}$ cm$^{-2}$ (Borkowski, Balbus & Fristrom 1990; Slavin 1989). The value is not expected to depend too much on the metallicity of the gas. As the metallicity is reduced, the cooling time of the gas in the interface increases which approximately compensates for the lowering of the metallicity in the calculation of the expected amount of O VI. With a total O VI column density of 2.3x10$^{14}$ cm$^{-2}$, ~23 cool/hot gas interfaces are required. Such a large number of interfaces is incompatible with the kinematical simplicity of the absorption produced by H I and the moderate states of ionization. However, turbulence in the gas could increase the number of interfaces and produce relatively simple low ionization line profiles.

Heckman et al. (2002) have followed the non-equilibrium cooling of various species for gas radiatively cooling from 5x10$^6$ K behind a 600 km s$^{-1}$ shock corresponding to a post shock flow velocity of 150 km s$^{-1}$. Their cooling flow calculations for isobaric and isochoric (constant volume) cooling and solar abundances are compared to the total observed column densities of O VI, O IV, Ne VIII, and S VI in Table 7. The assumed abundances are from Anders & Grevesse (1989) and the numerical values are listed in Table 5.

We use total observed column densities integrated through the entire absorber to compare with the theory. Different ions are expected to occur at different line of sight velocities in the flow depending on when they peak in abundance in the cooling flow. S VI was not included in the results tabulated by Heckman et al. (2002). D. Strickland, who performed the cooling flow calculations in the Heckman et al. publication, has kindly computed the values for S VI listed in Table 7.

The predicted abundances in cooling flow calculations are not particularly sensitive to the absolute metallicity in the gas so long as [Z/H] > -0.7, because as the metallicity is decreased, the cooling time increases. The increase in the cooling time compensates for the reduced metallicity when calculating logN(ion) through the cooling flow. For discussions of these compensating effects see Edgar & Chevalier (1996) and Savage et al. (2003). The absolute column densities in the cooling flows will scale with the flow velocity. Increases (or decreases) in the value of this velocity will produce increases (or decreases) in the absolute values of the predicted column densities. From Table 7 we see that in order to match the total observed O VI column density the flow velocity in the isobaric case must be increased by 0.1 dex while it must be decreased by 0.6 dex in the isochoric case.

The observed column density ratio of Ne VIII to O VI is most compatible with the results listed in Table 7 for isochoric cooling. Through the entire absorber we observe N(Ne VIII)/N(O VI) = 0.33±0.10, while the isochoric and isobaric cooling models predict ratios of 0.50 and 1.26, respectively. However, the large required decrease (0.6 dex) in the flow velocity to match the total column densities, may invalidate our use of the isochoric model calculation. We can therefore only conclude that the observations are inconsistent with isobaric cooling in a flow with a postshock velocity of ~150 km s$^{-1}$.

To properly explore cooling flow model predictions for the highly ionized species in the z = 0.20701 system toward HE 0226-4110 will require a grid of cooling flow calculations for various initial hot gas temperatures (or shock velocities) and for various assumed elemental abundances ranging from solar to 0.01 solar.

### 5.3. O VI, Ne VIII, and Searches for the WHIM

Although we do not know where the hot gas in the z = 0.20701 absorption line system is located (galaxy group, galaxy halo, or IGM filament), the observations imply the presence of large amounts of hot gas in the absorption system. These results, therefore, give support to the idea that at low redshift a substantial reservoir of baryons probably exists in a gas phase that up until now has been extremely difficulty to observe because it is so highly ionized. In the case of the hot gas absorber toward HE 0226-4110, we find an enormous ionization correction of N(H) = N(H I)x1.7x10$^6$ in the Ne VIII bearing gas assuming CIE.

Although the emphasis for searches for the WHIM have been on O VII systems, O VI systems, and broad Lyα lines, it is clear that Ne VIII is an important new diagnostic of gas in the temperature range from ~5x10$^5$ to ~1x10$^6$ K. As we have seen in this paper, the detection of Ne VIII can provide important information about the origin(s) of the ionization of the gas and can be used to discriminate between collisional ionization in hot gas and photoionization in cooler gas.

The primary difficulty in using Ne VIII absorption as an IGM diagnostic is obtaining adequate S/N at far-UV wavelengths where the relatively weak lines are expected to occur. It appears the earlier search for Ne VIII by Richter et al. (2004) toward PG 1259+593 may have come very close to detecting the stronger member of the doublet. We compare in Table 8 the properties of the z = 0.20701 system toward HE 0226-4110 to those O VI systems where strong upper limits have been placed on Ne VIII by Richter et al. (2004). In the four O VI systems toward PG 1259+593, N(Ne VIII)/N(O VI) ranges from $\leq$0.58(3σ) to $\leq$ 1.3(3σ). Except for hot gas in two narrow regions of T near 1.4x10$^6$ K and 2.6x10$^6$ K, these limits imply that none of the four O VI systems toward PG 1259+593 are tracing gas with T > 5.5x10$^5$ K, if the gas has a solar ratio of O to Ne and CIE conditions (see Fig. 7). This is an interesting result and may imply that hot gas systems with T > 5.5x10$^6$ K containing O VI are relatively rare or alternately, hot gaseous systems undergoing isobaric cooling are rare since N(Ne VIII)/N(O VI) = 1.26 in that case.

With a total hydrogen column density estimated to be N(H)= 8.3x10$^{19}$ cm$^{-1}$ for CIE and [Z/H]= -0.5, the column density of the z = 0.20701 system is ~5 times larger than for the two O VII systems detected at z = 0.011 and 0.027 in the X-ray spectrum of Mrk 421 (Nicastro et al. 2005). Since the highly favorable system we have analyzed has a Ne VIII rest wavelength equivalent width of only 32.9±10.5 mÅ for the stronger member of the doublet, it is evident that a careful study of the implications of the low redshift O VI/Ne VIII systems for determining the baryonic content of the low redshift IGM will require FUV observations with 3σ equivalent width detection sensitivities of ∼ 15 mÅ.   Although one of the limits for the Ne VIII λ770.409 line in Table 8 is this small, the three other 3σ limits have an average sensitivity of only 37 mÅ.

A simple test for the basic validity of the interpretations provided in this paper would follow from high quality measurements of the C IV λλ1548.204, 1550.781 doublet shifted to wavelengths near 1871 Å.   According to Figure 5 and Table 7, most of the C IV in the z = 0.20701 system should be associated with the photoionized absorber containing C III, N III, O III, Si III, O IV, and S VI.   C IV should have a column density of logN(C IV) = 14.0 and a velocity of −15 km s$^{-1}$. Measures of C IV would yield a more accurate measurement of [C/H] than is possible from the saturated C III absorption line.

The observations reported here illustrate the substantial diagnostic power of combined UV and FUV measurements to study the properties of gaseous structures in the low redshift IGM.

## 6. SUMMARY

High resolution FUSE and STIS observations of the bright QSO HE 0226-4110 ($z_{em}$ =0.495) covering the wavelength range from 916 to 1709 are presented for an IGM absorption line system at z (O VI) = 0.20701 containing H I (Ly α to κ), C III, O III, O IV, O VI, N III, Ne VIII, Si III, and S VI.  S V is tentatively detected at the 2.2σ level.  Significant non-detections are reported for C II, N II, N V, O I, O II, Si II, Si IV, and S IV. The detection significance of O VI, Ne VIII, and S VI absorption ranges  from 16σ for the  O VI doublet absorption,  to 3.9σ  for the Ne VIII doublet absorption,  to 3.3σ  for the stronger member of the S VI doublet.  These observations provide the first detection  of Ne VIII  in the IGM.

The absorption system is analyzed in order to obtain information about elemental abundances and physical conditions in the multi-phase absorption system.
1.   The H I (Ly α to κ) absorption is described by two absorbing components with v = -24±1 ±4 km s$^{-1}$, b = 17.4±1.4(+3.6,-2.3) km s$^{-1}$, and logN(H I) = 15.06±0.04 (+0.10,-0.05) for the principal absorbing component,  and v = 5 ±3 (+9,-2) km s$^{-1}$, b = 35.9±1.1 (+0, -2.2) km s$^{-1}$, and logN(H I) = 14.89±0.05(+0.11, -0.20)  for the second component.
2. The absorption by C III, O III, N III, Si III, O IV, and S VI near v = -15 km s$^{-1}$  is likely associated with the stronger H I absorbing component.  These intermediate and high ionization absorbers probably occur in a gas photoionized by the extragalactic background radiation with Solar heavy element ratios and an overall metallicity [Z/H] =-0.50±0.20.   Simple photoionization modeling reveals that the gas in the absorbing structure has T ∼ 2.1x10$^4$ K,  $n_H$ ∼  2.6x10$^{-5}$ cm$^{-3}$,  P/k ∼ 0.5  cm$^{-3}$ K,  N(H I) ∼1.15x10$^{15}$ cm$^{-2}$, N(H) ∼ 4.6x10$^{18}$ cm$^{-2}$,  N(H I)/N(H) ∼ 2.5x10$^{-4}$, and an absorber path length of ∼60 kpc.    This absorber may be tracing the warm photoionized gas in a galaxy group or gas in the outer most regions of a galaxy halo.
3.  Single component fits to the Ne VIII and O VI doublet absorption yields logN(Ne VIII) =13.89±0.11, b = 22.6±15 km s$^{-1}$, v = -7±6 km s$^{-1}$ and logN(O VI) =

14.37±0.03, b = 31.4±2.0 km s$^{-1}$, v = 0±2 km s$^{-1}$. Through the entire absorber N(Ne VIII)/N(O VI) = 0.33±0.10.

4. The ionization of O VI and Ne VIII is unlikely to be the result of photoionization by the extragalactic background because very low gas densities and long absorption path lengths are required to reproduce the high level of ionization. The photoionization origin is ruled out because the Hubble flow absorption line broadening over the long path length exceeds the observed breadth of the high ionization absorption lines by a factor of ~10.

5. The O VI and Ne VIII observations are likely revealing hot collisionally ionized gas. Under conditions of CIE, N(Ne VIII)/N(O VI) = 0.33±0.10 is achieved in gas with a solar ratio of O to Ne near T = 5.4x10$^5$ K. If the overall metallicity in the gas is [Z/H] = -0.50, the observed column densities of Ne VIII and O VI imply logN(H I) ~13.69 and logN(H) ~ 19.9 cm$^{-2}$ in the gas containing Ne VIII. The value of N(H I) is compatible with the possible presence of a broad component to the Ly α absorption.

6. The observations of O VI and Ne VIII in the z = 0.20701 system support the idea that a substantial fraction of the baryonic matter at low redshift exists in hot gaseous structures that up until now have been very difficult to study because of the extremely high levels of ionization. In the system studied here N(H)/N(H I) ~1.7x10$^6$ in the Ne VIII bearing gas assuming CIE.

7. The observations reveal the considerable power of UV, far-UV, and EUV absorption line diagnostics for gaining insights about elemental abundances and physical conditions in the multi-phase gas of the IGM.

We thank the HST and FUSE mission operations teams for their efforts to provide excellent UV and far-UV spectroscopic capabilities to the astronomical community. We thank D. Strickland of JHU for providing the cooling flow calculations for S VI that are listed in Table 7. We thank J. M. Shull and M. Giroux for their comments about an early draft version of this paper. An anonymous referee provided a number of important suggestions that helped us to improve our paper during the review process. Our work is based on data obtained by the NASA-CNES-CSA FUSE mission operated by the Johns Hopkins University. The STIS observations were obtained as part of HST program 9184, with support through NASA grant HST-GO-9184.08-A from the Space Telescope Science Institute. BDS acknowledges support through NASA Grant NNG-04GC70G. BPW acknowledges support from NASA grants NAG5-9024 and NAG5-9179. Partial financial support has been provided to KRS by NASA contract NAS5-32985 and to TMT by NASA through Long Term Space Astrophysics grant NNG-04GG73G. Some of the data presented in this paper were obtained from the Multimission Archive at the Space Telescope Science Institute (MAST), STScI is operated by the Association of Universities for Research in Astronomy Inc. , under NASA contract NAS5-26555.

REFERENCES

Allende Prieto, C., Lambert, D. L., & Asplund, M. 2002, ApJ, 573, L137

Anders, E., & Grevesse, N. 1989, Geochim. Cosmochim. Acta, 53, 197

Asplund, M., Grevesse, N., Sauval, A. J., Allende Pieto, C., & Kiselman, D. 2004, A& A, 417, 751

Borkowski, K. J., Balbus, S. A., & Fristrom, C. C. 1990, ApJ, 355, 501

Cen, R., & Ostriker, J. P. 1999, ApJ, 514, 1

Collins, J. A., Shull, J. M., & Giroux, M. L. 2004, ApJ, 605, 216

Davé, R., et al. 2001, ApJ, 552, 473

Edgar, R., J., & Chevalier, R. A. 1986, ApJ, 310, L27

Fang, T., Marshall, H. L. Lee, J. C., Davis, D. S., & Canizares, C. R. 2002, ApJ, 572, L127

Fang, T., Sembach, K. R., & Canizares, C. R. 2003, ApJ, 586, L49

Ferland, G. J., Korista, K. T., Verner, D. A., Ferguson, J. W., Kingdon, J. B., & Verner, E. M. 1998, PASP, 110, 761

Fitzpatrick, E. L., & Spitzer, L. 1997, ApJ, 475, 623

Futamoto, K, Mitsuda, K., Takei, Y., Fujimoto, R., & Yamasaki, N. Y. 2004, ApJ, 605, 793

Ganguly, R., Masiero, J., Charlton, J. C., & Sembach, K.R. 2003, ApJ, 598, 922

Grevesse, N., & Sauval, A. J. 1998, Space Science Reviews, 85, 161

Haardt, F., & Madau , P. 1996, ApJ, 461, 20

Heckman, T. M., Norman, C. A., Strickland, D. K., & Sembach, K. R. 2002, ApJ, 577, 691

Holweger, H. 2001, in Solar and Galactic Composition, AIP Conf. Proc. 598, ed. R. F. Wimmer-Schweingruber, 23 (astro-ph/0107426)

Indebetouw, R., & Shull, J. M. 2004, ApJ, 605, 2051

Kimble, R. A., et al. 1998, ApJ, 492, L83

Mathur, S., Weinberg, D. H., & Chen, X. 2003, ApJ 582, 82

McKernan, B., Yaqoob, T., & Reynolds, C. S. 2004, ApJ, 617, 232

Moos, H.W. et al. 2000, ApJ, 538, L1

Morton, D. 2003, ApJS, ApJS, 149, 205

Nicastro, F. et al. 2002, ApJ, 573, 157

Nicastro, F. et al. 2005, Nature, 433, 495

Paerels, F. B. S. & Kahn, S 2003, ARAA, 41, 291

Prochaska, J. X., Chen, H.-W., Howk, J.C., Weiner, B. J., & Mulchaey, J. 2004, ApJ, 617, 718

Proffitt, C. et al. 2002, STIS Instrument Handbook, v6.0, (Baltimore:STScI)

Rasmussen, A., Kahn, S. M., & Paerels, F. 2003, in The IGM/Galaxy Connection, ed. J. Rosenberg & M. E.Putman, (Dordrecht: Kluwer), 109

Richter, P., Savage, B. D., Tripp, T. M., & Sembach, K. R. 2004, ApJS, 153, 165

Savage, B. D., & Sembach, K. R. 1991, ApJ, 379, 245

Savage, B. D., Sembach, K. R., Tripp, T. M., & Richter, P. 2002, ApJ, 564, 631

Savage, B. D., Sembach, K. R., Wakker, B. P., Richter, P., Meade, M., Jenkins, E. B., Shull, J. M., Moos, H. W., & Sonneborn, G. 2003, 146, 165

Savage, B. D., Wakker, B. P., Fox, A., & Sembach, K. R. 2005, ApJ, 619, 863

Sembach, K. R., & Savage, B. D. 1992, ApJS, 83, 147

Sembach, K. R., Tripp, T. M., Savage, B. D., & Richter, P. 2004, ApJS, 155, 351

Sahnow, D. S., Moos, H. W., Ake, T., et al. 2000, ApJ, 537, L7

Shull, J. M., Tumlinson, J, & Giroux, M. L. 2003, ApJ, 594, L107

Slavin, J. D. 1989, ApJ, 346, 718

Sutherland, R. S., & Dopita, M. A. 1993, ApJS, 88, 253

Tripp, T. M., Savage, B. D., & Jenkins, E. B. 2000, ApJ 534, L1

Tripp, T. M., Giroux, M. L., Stocke, J.T., Tumlinson, J., & Oegerle, W. R. 2001, ApJ, 563, 724

Tripp, T. M., Jenkins, E. B., Williger, G. M., Heap, S. R., Bowers, C. W., Danks, A. et al. 2002, ApJ, 575, 697

Tripp, T. M., Jenkins, E. B., Bowen, D. V., Prochaska, J. X., Aracil, B., & Ganguly, R. 2005, ApJ, 619, 714

Tumlinson, J., Shull, J. M., Giroux, M. L., & Stocke, J. T. 2005, ApJ, 620, 95

Verner, D. A., Verner, E. M., & Ferland, G. J. 1996, Atomic Data Nucl. Data Tables, 64, 1

Wakker, B. P., Savage, B. D., Sembach, K. R., Richter, P., Meade, M., Jenkins, E. B. et al. 2003, ApJS, 146, 165

Woodgate, B. et al. 1998, PASP, 110, 1183

Yao, Y., & Wang, Q. D. 2005, ApJ, 625, June 1 issue, (astro-ph/0502242)

Yuan, Q., Green, R. F., Brotherton, M., Tripp, T. M., Kaiser, M. E., & Kriss, G. A. 2002, ApJ, 575, 687

TABLE 1
LOG OF FUSE AND STIS OBSERVATIONS OF HE 0226-4110[a]

| Instrument/data set ID | Obs. date | Exp. Time (ks) |
|---|---|---|
| FUSE/P2071301 | 2000-Dec-12 | 11.0 |
| FUSE/P1019101 | 2001-Oct-03 | 50.3 |
| FUSE/P1019102 | 2002-Nov-15 | 14.5 |
| FUSE/P1019103 | 2002-Nov-16 | 18.9 |
| FUSE/P1019104 | 2002-Nov-17 | 18.1 |
| FUSE/D0270101 | 2003-Sept-01 | 23.9 |
| FUSE/D0270102 | 2003-Sept-03 | 39.9 |
| FUSE/D0270103 | 2003-Oct-31 | 16.7 |
|  |  | Total=193.3 |
|  |  |  |
| STIS/O6E107010,20,30 | 2002-Dec-26 | 8.1 |
| STIS/O6E108010,20,30 | 2002-Dec-26 | 8.1 |
| STIS/O6E109010,20,30 | 2002-Dec-27 | 8.1 |
| STIS/O6E110010,20,30 | 2002-Dec-29 | 8.1 |
| STIS/O6E111010,20 | 2002-Dec-31 | 5.1 |
| STIS/O6E111030,40 | 2003-Jan-01 | 6.0 |
|  |  | Total=43.5 |

[a] The FUSE observations were obtained in the time tagged mode using the 30"x30" aperture. The STIS observations were obtained with the E140M grating set for a central wavelength of 1425Å.

TABLE 2
METAL LINE MEASUREMENTS[a]

| ION | $\lambda_{rest}$ (Å) | S/N | $\langle v \rangle \pm \sigma$ (km s$^{-1}$) | $b \pm \sigma$ (km s$^{-1}$) | $W_{rest} \pm \sigma$ | $\mathrm{Log}N_a(\mathrm{cm}^{-2}) \pm \sigma$ (dex) | Grating/ Segment | Note |
|---|---|---|---|---|---|---|---|---|
| C II | 903.962 | 17 | … | … | <22.6 (3σ) | <12.97(3σ) | LiF2A | 1 |
| C II | 1334.532 | 6 | … | … | <41.8 (3σ) | <13.32(3σ) | E140M | |
| C III | 977.020 | 14 | -9 ±4 | 27.2±6.2 | 216±14.2 | >13.90 | LiF2A | 2 |
| N II | 1083.994 | 11 | … | … | <27.6(3σ) | <13.36(3σ) | E140M | |
| N III | 989.799 | 5 | -15 ±8 | 16.4: | 28.2±11.5 | 13.56(+0.13,-0.30) | E140M | 3 |
| N V | 1242.804 | 10 | … | … | <31.1(3σ) | <13.46(3σ) | E140M | 4 |
| O I | 1302.169 | 11 | … | … | <22.5(3σ) | <13.46(3σ) | E140M | |
| O II | 834.465 | 13 | … | … | <27.6(3σ) | <13.54(3σ) | LiF1A | |
| O III | 832.929 | 17 | -10±3 | 23.1±3.1 | 90.8±12.4 | 14.30±0.06 | LiF1A | |
| O IV | 787.711 | 8 | -10±5: | >17.4 | >91.1 | >14.50: | SiC2A | 5 |
| O VI | 1031.926 | 10 | 0±2 | 29±2.1 | 169±15.3 | 14.36±0.05 | E140M | 6 |
| O VI | 1037.617 | 11 | -2±3 | 31.3±1.7 | 112±10.4 | 14.38±0.04 | E140M | 6 |
| Ne VIII | 770.409 | 11 | -3±8 | 28.3±8.8 | 32.9±10.5 | 13.85(+0.12,-0.17) | SiC2A | 7 |
| Ne VIII | 780.324 | 9 | -18±13 | 18.9: | 24.9±10.6 | 14.03(+0.15,-0.24) | SiC2A | 7 |
| Si II | 1260.422 | 9 | … | … | <26.7(3σ) | <12.21(3σ) | E140M | |
| Si III | 1206.500 | 10 | -15±2 | 14.8±2.1 | 44.2±8.1 | 12.40±0.08 | E140M | |
| Si IV | 1402.773 | 4 | … | … | <63.7(3σ) | <13.16(3σ) | E140M | 8 |
| S IV | 1062.664 | 10 | … | … | <24.9(3σ) | <13.70(3σ) | E140M | |
| S V | 786.468 | 8 | -14±7 | 19.0: | 22.7±10.2 | 12.49(+0.16,-0.28) | SiC2A | 9 |
| S VI | 933.378 | 17 | -20±9 | 37.5±6.8 | 27.0±8.2 | 12.93(+0.12,-0.17) | LiF2A | |
| S VI | 944.523 | 22 | … | … | <16.6(3σ) | <12.99(3σ) | LiF2A | |

[a] The apparent optical depth (AOD) method of Savage & Sembach (1991) was used to determine the apparent column densities, $\log N_a$. The integration ranges for the detected ions are indicated above the absorption lines shown in Figures 2a and b by the horizontal lines. For very strong absorption lines of C III and O IV the AOD column density is listed as a lower limit. When a column density upper limit is listed, the 3σ limit is derived from the 3σ equivalent width limit and the assumption the absorption is on the linear part of the curve of growth. The 3σ limits were derived for absorption lines 50 km s$^{-1}$ wide. Values of λ and the f-value are taken from Morton (2003) for λ > 912 Å and from Verner et al. (1996) for λ < 912 Å.

Notes: (1) We searched for the C II λ903.962 line with f= 0.336. The weaker C II λ903.624 line with f = 0.168 has been neglected in the analysis. (2) The C III λ977.020 absorption is so strong the AOD column density is listed as a lower limit. (3) Blending with the Si II λ989.873 line is not a problem because the much stronger Si II λ1260.422 line is not detected. (4) The 3σ limits for $W_{rest}$ and logN are given for N V based on the λ1242.804 line. The stronger N V λ1238.821 line is badly blended with a Ly α line at z = 0.23010 (see Fig. 2). (5) The O IV λ787.711 absorption blends with ISM O I λ950.885 (see Fig. 2). The measurements refer to the O IV absorption over the velocity range from –60 to 0 km s$^{-1}$. Because of the blending and the great depth of the O IV absorption, the column density is listed as a lower limit. (6) The O VI measurements are integrated over the velocity range from –55 to 55 km s$^{-1}$. (7) The combined FUSE observations for the SiC2A segment near 930 Å required and extra 4 km s$^{-1}$ zero point velocity shift in order to properly align the near by ISM lines of O I and D I. The Ne VIII observations are integrated over the velocity range from –50 to 42 km s$^{-1}$. If the two Ne VIII lines are instead integrated over the range from –52 to 20 km s$^{-1}$, $W_{rest}$ = 33.0±10.3 mÅ and 26.7±9.2 mÅ, for the strong and weak line, respectively. (8) The 3σ limits for $W_{rest}$ and logN are given for Si IV based on the

$\lambda 1402.773$ line. The stronger Si IV $\lambda 1393.760$ line is badly blended with a Ly $\alpha$ line at z = 0.38427 (see Fig. 2). (9) The S V $\lambda 786.468$ absorption is a tentative 2.2σ detection. A 3σ upper limit for the absorption is $W_{rest} < 30.6$ mÅ and logN(S V) <12.6. The S V absorption is bracketed by terrestrial airglow emission from O I $\lambda 948.685$ and H I $\lambda 949.743$. Inspection of the night only data reveals that this emission is not affecting the S V measurement.

TABLE 3
ABSORPTION LINE PROFILE FIT RESULTS[a]

| Ion | $\lambda$ (Å) | $\chi_v^2$ | v±σ (km s$^{-1}$) | b±σ (km s$^{-1}$) | LogN(cm$^{-2}$)±σ (dex) | Note |
|---|---|---|---|---|---|---|
| H I | Lyman lines | 0.906 | -24±1 | 17.4±1.4 | 15.06±0.04 | 1 |
| " | " | | 5±2 | 35.9±1.1 | 14.89±0.05 | |
| H I | Lyman lines | 0.802 | -25 ±3 | 14.3±3.9 | 15.01±0.21 | 2 |
| " | " | | 5±2 | 38.5±1.3 | 14.79±0.06 | |
| C III | 977. 020 | 0.446 | -15 | 17.2±3.3 | 14.04±0.22 | 3 |
| " | " | | 10 | 22.3±5.0 | 13.20±0.07 | |
| N III | 989.799 | 0.881 | -12 ±6 | 16.7: | 13.61±0.16 | |
| O III | 832.929 | 1.13 | -15 | 15.1±4.2 | 14.23±0.07 | 4 |
| " | " | | 10 | 16.9: | 13.66±0.16: | |
| Si III | 1206.500 | 0.998 | -15±3 | 17.4±5.5 | 12.46±0.08 | 5 |
| O VI | 1031.926, 1037.617 | 1.04 | 0±2 | 31.4±2.0 | 14.37±0.03 | 6 |
| Ne VIII | 770.409, 780.324 | 0.601 | -7±6 | 22.6±15.0 | 13.89±0.11 | |
| S VI | 933.378, 944.523 | 2.73 | -18 ±9 | 38.8±19.9 | 12.78±0.11 | 7 |

[a] The component fit errors listed here only allow for statistical errors in the component fit process. These errors do not include the sometimes dominant systematic errors associated with the uncertainty of the simple kinematical assumptions. In this case we assume that the absorption is reliably described by two simple Voigt optical depth profiles. Table 4 was prepared to explore the sensitivity of the component fit results to component parameters for H I. In §3 we discuss the systematic errors on the fit parameters for H I associated with separately determining the properties of the two H I absorbers.

Notes: (1) The H I component fit results listed here are based on a simultaneous two-component fit to the observed profiles for Ly $\alpha$ to Ly $\theta$. The profiles simultaneously fitted include Ly $\alpha$, Ly $\beta$, and Ly$\gamma$ profiles from STIS and the Ly $\gamma$ to Ly $\theta$ profiles from the FUSE LiF2A channel observations. The fit also includes the FUSE LiF1B channel H I observations for Ly $\gamma$, Ly $\delta$, and Ly $\epsilon$. (2) This H I component fit result is a free two-component fit to only the STIS Ly $\alpha$, Ly $\beta$, and Ly $\gamma$ profiles. The large error for logN for the principal component obtained in this fit is the result of only fitting the very strong absorption lines. (3) The component velocities for C III were fixed when performing the fit. The two-component fit was used to better constrain the column density of C III in the strong component. (4) The component velocities of O III were fixed when performing the fit. The values for the weaker component are very uncertain. The two-component fit was used to better constrain the column density of the strong component. (5) The reliably detected Si III absorption provides the best information on the velocity of the intermediate ionization absorption component traced by C III, N III, O III, and Si III. (6) The O VI results listed here are for a single component fit to the observations. (7) The detected and non-detected lines of the S VI doublet were simultaneously fitted. The inclusion of the non-detected $\lambda 944.523$ line explains why the value of logN(S VI) from the fit process is 0.15 dex smaller that the value obtained from the $N_a$(v) integration listed in Table 2.

TABLE 4
LACK OF SENSITIVITY OF  H I COMPONENT FITTING TO
COMPONENT PARAMETERS[a]

| Case | Component | v±σ (km s$^{-1}$) | b±σ (km s$^{-1}$) | LogN(cm$^{-2}$)±σ (dex) | $\chi_v^2$ |
|------|-----------|------|------|----------|------|
| 1 | 1 | -30 | 13.0±1.1 | 14.94±0.03 | 0.954 |
| " | 2 | 2±1 | 34.7±0.7 | 15.02±0.02 | |
| 2 | 1 | -27 | 15.1±1.0 | 15.01±0.03 | 0.917 |
| " | 2 | 3±2 | 35.3±0.8 | 14.96±0.03 | |
| Best fit | 1 | -24 | 17.4±1.0 | 15.06±0.03 | 0.904 |
| " | 2 | 5±2 | 35.9±1.1 | 14.89±0.04 | |
| 3 | 1 | -20 | 21.0±0.6 | 15.16±0.04 | 0.922 |
| " | 2 | 14±6 | 33.7±3.9 | 14.69±0.09 | |
| 4 | 1 | -17 | 22.9±0.5 | 15.21±0.02 | 0.951 |
| " | 2 | 26±4 | 28.4±3.0 | 14.46±0.06 | |
| 5 | 1 | -15 | 24.3±0.5 | 15.23±0.02 | 0.979 |
| " | 2 | 32±4 | 26.1±3.1 | 14.32±0.07 | |

[a] Simultaneous two component fits to the STIS and FUSE Lyman series absorption in the z = 0.20701 O VI system.   The velocity of the principal component (#1) was fixed and the other component parameters were allowed to vary to obtain the best fit solution.  The smallest value of  $\chi_v^2$  is obtained for a principal component velocity of –24 km s$^{-1}$  as reported for the fully free fit result listed in Table 3.  However, since the two absorbers have roughly similar b values and column densities,  the minimum of  $\chi_v^2$  has a broad distribution as a function of the velocity of the principal absorption component.  The fit for a principal component velocity of –24 is hardly better than for –17 or –30 km s$^{1}$.  The uncertainties associated with the fitting process are further compounded by our lack of knowledge of the true component structure of the absorber.  For example, adding in additional weak component or asymmetries to the principal components would substantially modify the fit parameters for the principal components.

TABLE 5
RECENT SOLAR ABUNDANCE REVISIONS

| Element | log(X/H)$_O$ (Anders & Grevesse 1989) | log(X/H)$_O$ (Grevesse & Sauval 1998) | log(X/H)$_O$ (Revision) | Reference |
|---------|------|------|------|-----------|
| C | -3.44 | -3.48 | -3.61 | Allende Prieto et al. (2002) |
| N | -3.95 | -4.08 | -4.07 | Holweger (2001) |
| O | -3.07 | -3.17 | -3.34[a] | Asplund et al. (2004) |
| Ne | -3.91 | -3.92 | -4.16[a] | Asplund et al. (2004) |
| Mg | -4.41 | -4.42 | | |
| Si | -4.45 | -4.45 | -4.46 | Holweger (2001) |
| S | -4.79 | -4.80[b] | | |
| Fe | -4.49 | -4.50 | | |

[a] Over the period from 1989 to 2004 the solar oxygen abundance has decreased  by 0.27 dex or a factor of 1.86.  The decrease for Neon has been 0.25 dex or a factor of 1.78.

[b] Meteoric value for log(S/H)  listed here is 0.13 dex smaller than the uncertain photospheric value.

TABLE 6
PROPERTIES OF THE PHOTOIONIZED ABSORBER
CONTAINING  C III, O III, N III, Si III, O IV, and S VI

| [Z/H] | -0.50 ±0.20 |
|---|---|
| log U | -1.85 |
| T (K) | $2.1 \times 10^4$ |
| $n_H$(cm$^{-3}$) | $2.6 \times 10^{-5}$ |
| P/k (cm$^{-3}$K) | 0.54 |
| N(H I) (cm$^{-2}$) | $1.15 \times 10^{15}$ |
| N(H) (cm$^{-2}$) | $4.6 \times 10^{18}$ |
| N(H I)/N(H) | $2.5 \times 10^{-4}$ |
| L(kpc) | 57 kpc |

[a] The observations are consistent with Solar abundance ratios
for N, O, Si, and S  to within 0.15 dex.  The error is larger for C
based on the saturated measurements of C III.  A discussion of
the errors for  the model fit parameters is in  §4.

TABLE  7
COLUMN DENSITES AND IONIC RATIOS OF COOLING IONS[a]

| Quantity | Isobaric | Isochoric | Observed[b] |
|---|---|---|---|
| logN(Ne VIII) | 14.3 | 14.7 | 13.89±0.11 |
| logN(O VI) | 14.2 | 15.0 | 14.37±0.03 |
| logN(O IV) | 13.6 | 15.0 | >14.50[c] |
| logN(N V) | 12.6 | 13.6 | <13.54 |
| logN(C IV) | 12.5 | 13.9 | … |
| logN(S VI) | 11.4[d] | 12.6[d] | 12.78±0.11[d] |
| N(Ne VIII)/N(O VI) | 1.26 | 0.50 | 0.33±0.10 |
| N(O IV)/ N(O VI) | 0.25 | 1.00 | >1.3[c] |
| N(N V)/ N(O VI) | 0.025 | 0.04 | <0.15 |
| N(S VI)/ N(O VI) | 0.0016 | 0.004 | 0.026±0.008 |

[a] The cooling flow calculations of Heckman et al. (2002) are listed for isobaric and isochoric cooling
columns and the solar abundances listed in Table 5 from  Anders & Grevesse (1989).
Since S VI was not included in the original Heckman et al. paper, D. Strickland kindly provided us with
the values logN(S VI) = 11.4 and 12.6 for isobaric and isochoric cooling behind a 600 km s$^{-1}$shock.  The
model predictions for the various ions should not change significantly as the metallicity in the cooling
gas is changed from solar to 0.3 solar. As the metals are removed from the gas the cooling times
lengthen.  This effect approximately compensates for   the lowering of the metallicity when computing
values of N(ion). For more detailed discussions of this effect see Edgar & Chevalier (1986) and Savage
et al. (2003).
[b] We list observed column densities and ratios for absorption integrated through the entire absorber.  No
allowance has been made for the observed velocity offsets between O VI, Ne VIII, and other species.
[c] In the case of O IV we have listed the lower limit to the total amount of O IV observed in the
absorption system.  However, all of the observed O IV can be explained by photoionization in the
moderate ionization system containing C III, N III, O III, and Si III.
[d] We have listed the total amount of S VI observed in the absorption system.  However, all of the
observed S VI can be explained by photoionization in the moderate ionization system containing C III, N
III, O III, and Si III.

TABLE 8
SEARCHES FOR Ne VIII in QSO ABSORPTION LINE SYSTEMS CONTAINING O VI

| QSO | z | O VI λ1031.926 $W_{rest}$(mÅ) | Ne VIII λ770.409 $W_{rest}$(mÅ) | O VI logN(cm$^{-2}$) | Ne VIII logN(cm$^{-2}$) | N(Ne VIII)/ N(O VI) | Note |
|---|---|---|---|---|---|---|---|
| HE 0226-4110 | 0.20701 | 169±15.3 | 32.9±10.5 | 14.37±0.03 | 13.89±0.11 | 0.33±0.10 | 1 |
| PG 1259+593 | 0.21949 | 90±7 | <38 (3σ) | 13.96±0.04 | <13.94(3σ) | <0.95(3σ) | 2 |
| " | 0.22313 | 57±14[a] | <30 (3σ) | 13.99±0.11 | <13.75(3σ) | <0.58(3σ) | 2 |
| " | 0.25971 | 76±12 | <43(3σ) | 13.84±0.07 | <13.96(3σ) | <1.3σ) | 2 |
| " | 0.31978 | 25±4 | <15(3σ) | 13.44±0.06 | <13.57(3σ) | <1.3σ) | 2 |

[a] The rest equivalent width listed for O VI is for the λ1037.617 component of the doublet. The λ1031.926 line was blended.

Notes: (1) This is the Ne VIII detection from this paper. (2) These upper limits to Ne VIII in the O VI system absorbers are from Richter et al. (2004).

FIG. 1. – Flux (erg cm$^{-2}$ s$^{-1}$ Å$^{-1}$) versus heliocentric wavelength (Å) over the region from 1235 to 1255 Å for the STIS E140M spectrum of HE 0226-4110. The STIS spectrum illustrated has a resolution of 7 km s$^{-1}$ (FWHM) and is binned into pixels ~7 km s$^{-1}$ wide. The Ly β λ1025.722, and O VI λλ1031.926, 1037.617 absorption lines in the multiphase system at z = 0.20701 are identified. Various other IGM and ISM absorbers labeled 1-6 include: (1) Ly α at z= 0.01748, (2) O III λ832. 929 at z = 0.49253, (3) Ly α at z = 0.02681, (4) ISM S II λ1250.578, (5) Ly β at z = 0.22005, and (6) ISM S II λ1253.805.

FIG. 2. - Continuum normalized flux versus rest frame velocity for lines detected in the STIS and FUSE observations of the multi-phase z = 0.20701 system in the spectrum of HE 0226-4110 are shown in (a) and (b). Important non-detections are also shown. The STIS observations have been binned by a factor of 2 into pixels ~7 km s$^{-1}$ wide. The FUSE observations are binned a factor of 4 into pixels ~ 7.5 km s$^{-1}$ wide. The continuum levels are displayed as the horizontal dashed lines. Contamination from other interfering absorption lines is identified. The emission bracketing the ISM O I λ929 absorption in the Ne VIII λ770 panel is from terrestrial O I airglow emission. Absorption lines with λ$_{rest}$ > 980 Å are from STIS and observations with λ$_{rest}$ < 980 Å are from FUSE. The vertical dotted lines define the velocity range of the O VI λλ1031.926, 1037.617 absorption. The horizontal bars above the detected metal lines indicate the velocity range of the N$_a$(v) integrations listed in Table 2.

FIG. 3. - Flux versus rest frame velocity for the Ne VIII λλ 770.409, 780.324 doublet in the system at z = 0.20701. A larger velocity range is shown than for Figures 2b and 4 in order to better display the continuum fit including H$_2$ absorption (heavy solid lines) and the noise behavior of the observations over a large velocity range. In the two panels the night+day observations are shown with the black histogram. In the upper panel the night-only observations are also shown with the red histogram. The vertical dashed lines show the velocity extent of the Ne VIII absorption. The FUSE observations have been binned to 4 pixels corresponding to 7.5km s$^{-1}$. Various other absorption lines and airglow emission are indicated as in Figures 2b and 4. The wide velocity range in this figure reveals additional absorption by ISM H I λ930.748 along with corresponding H I airglow emission in the upper panel. The weak 1.5σ feature near 380 km s$^{-1}$ in the lower panel is unidentified and may not be real. The strongest feature in the lower panel is the 2.3σ feature we identify as Ne VIII near v = 0 km s$^{-1}$.

FIG. 4. - Continuum normalized flux versus rest frame velocity for selected absorption lines detected in the STIS and FUSE observations of HE 0226-4110 for the system at z = 0.20701. The FUSE spectra have been binned into 7.5 km s$^{-1}$ wide pixels. The STIS spectra are shown with the full 3.5 km s$^{-1}$ detector sampling. Voigt profile fits to the H I Lyman lines and to most of the metal lines listed in Table 3 are shown. The tic marks above the profiles denote absorption component velocities either required by the fit or adopted in the fitting process (see notes to Table 3). The dashed and dotted lines fitted to the HI δ λ949 absorption line show the fit results for model case #1 and #4 in Table 4.  Those fits are beginning to become distinguishable from the solid line best fit result.

FIG. 5. - Photoionization model  for the absorption lines of C III, N III, O III, Si III, O IV, and S VI seen in the spectrum of HE 0226-4110 at z = 0.20701.   The plane parallel slab model assumes the photoionizing extragalactic background at z = 0.2 from Haardt and Madau (1996), a uniform constant density slab  with a H I column density of logN(H I) = 15.06,  an overall  metallicity of [Z/H] = -0.50, and Solar abundance ratios among the heavy elements.  The various curves show the expected run of logN(ion) vs log U,  where U is the ionization parameter.  Those curves denoted with light solid lines are for species for which observational data exists.  No observations exist for C IV (denoted with the dashed line).  Various other quantities required by the model fit are also shown including  the  total H density, $n_H(cm^{-3})$, the total H column density,  N(H) = $N_H(cm^{-2})$, and the path length through the absorber L(kpc).  For all the detected ions the ranges of log U where the model agrees with the observed column densities to within ±1σ  are shown with the  heavy black lines attached to the model curves.  For C III, N III, O III,  Si III, O IV, and S VI the model and the  observed column densities  agree for log U = -1.85 (see the dashed vertical line).  This model also explains the tentative 2.2σ detection of S V.  However,  the photoionized model  for these species does not explain the observed amounts of O VI and Ne VIII which require the presence of collisionally ionized gas.

FIG. 6.- A photoionization model for the highly ionized absorption lines seen in the spectrum of HE 0226-4110 at z = 0.20701.  The model assumes with logN(H I) = 13.67 and [Z/H] = -0.5.   The various axes are the same as for Figure 5 except for the addition of a Hubble flow broadening axis where the broadening width  $\Delta v$ (km s$^{-1}$) = 70xL(Mpc).   The very large ionization parameter logU = -0.09 is required to fit the observed column densities of O VI and Ne VIII in the highly ionized absorber.   Such a model is ruled out because the very low density in the gas requires a  large path length of ~11 Mpc.   The Hubble flow over such a path would make the absorption lines ~10 times broader than the observed absorption.

FIG. 7 -  A  CIE   model for  O VI and Ne VIII in the  z = 0.20701 system toward HE 0226-4110.   The curves display logN$_{ion}$(cm$^{-2}$) for H I, O IV, O VI, Ne VIII,  S VI, and C IV as a function of logT(K).   The model assumes logN(H) = 19.92,  [Z/H] = -0.50, and a Solar abundance ratio among the heavy elements.  The observed O VI and  Ne VIII column densities  in the highly ionized gas absorber agree with the model  for log T =  5.73.  At this value of T, the model predicts much less O IV and S VI than observed.  The O IV and S VI appear to arise in the photoionized gas modeled in Figure 5.  No observations of C IV have been obtained in this absorption system.

FIG. 8. -  The histograms in the three panels show the STIS continuum normalized H I Ly α profile plotted against restframe velocity  for  z = 0.20701 absorber.  The dotted lines show the continuum levels.   The solid line in panel (a) shows the best fit two-component H I model using the results from  Table 3 for the fit to the STIS Ly α,

Ly β , and Ly γ  data  alone.  The vertical tick marks indicate the velocities of the components. The solid line in panel (b) shows the  best three-component model fit including the two components from (a) and an additional third broad component for which the best free fit results are found to be: logN(H I) = 13.31±0.39, v = -7 km s$^{-1}$, and b = 97 km s$^{-1}$.  The dashed line shows the broad component plotted separately while the solid line shows the total absorption.    The solid line in (c) shows a three component fit with the broad component parameters:   logN(H I) = 13.70, v = -7 km s$^{-1}$, and b = 97 km s$^{-1}$. The value of logN(H I) for the broad component in this case is taken to be 1σ larger than for the best fit  three-component result shown in (b).   The observed Ly α profile is compatible with the possible existence of a broad component having logN(H I) in the range from ~13.3 to ~13.7.

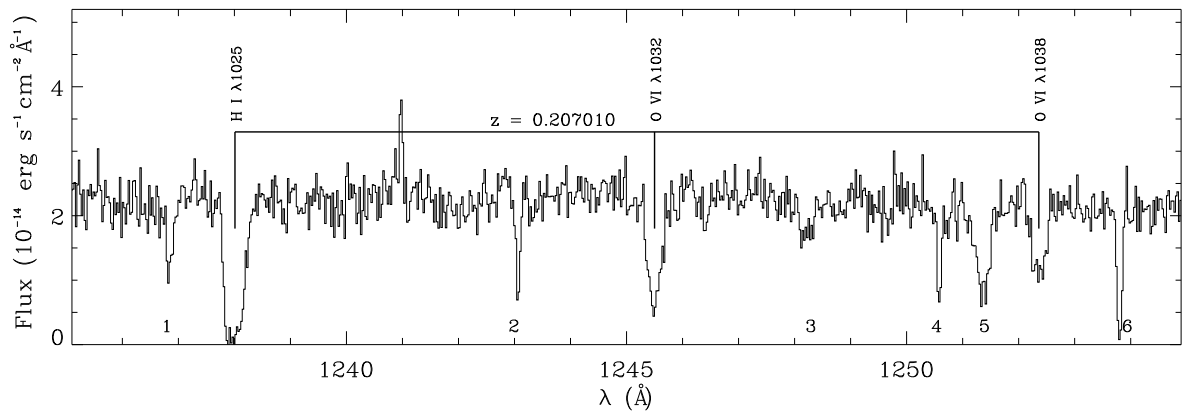

Fig. 1.—

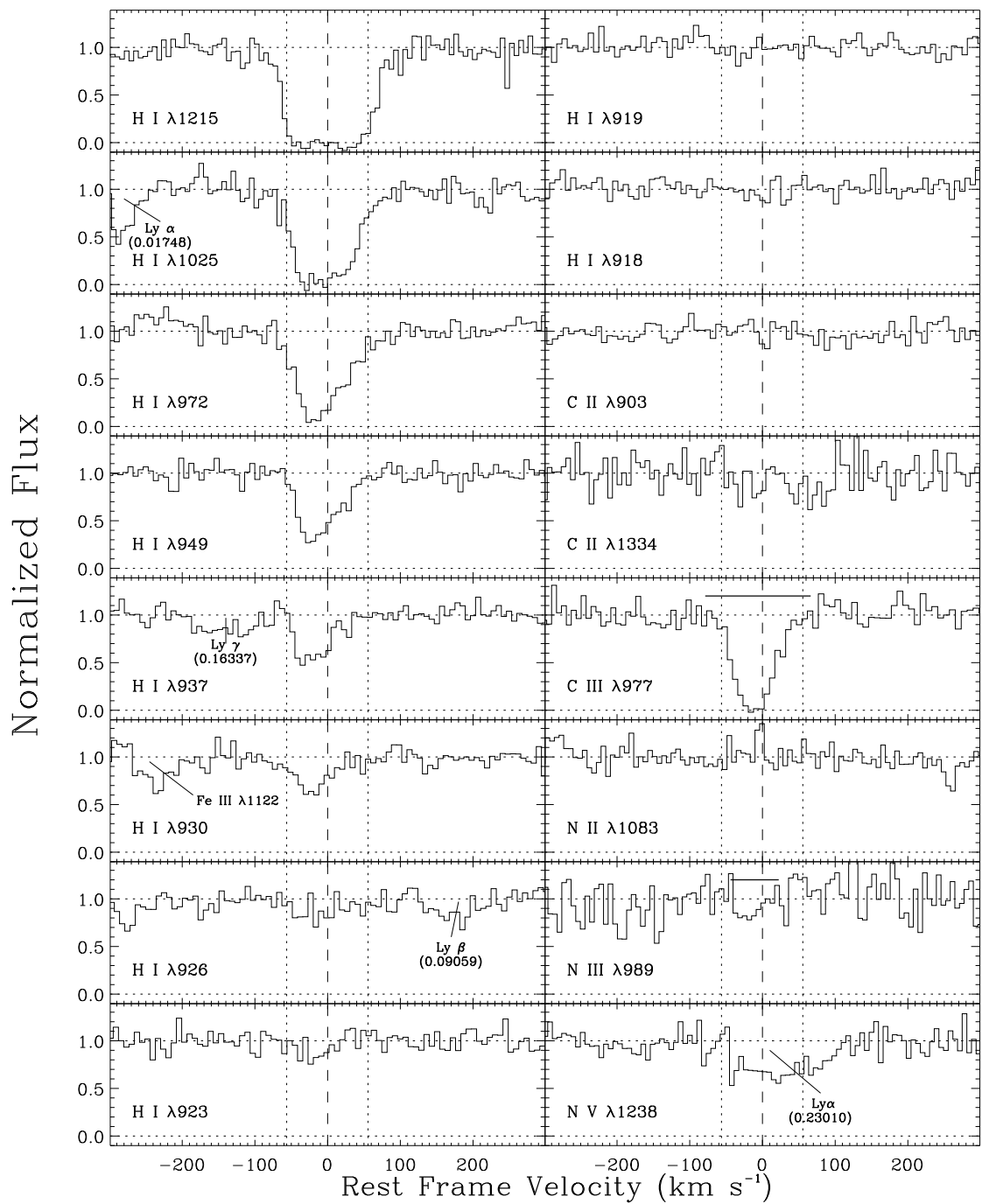

Fig. 2.—

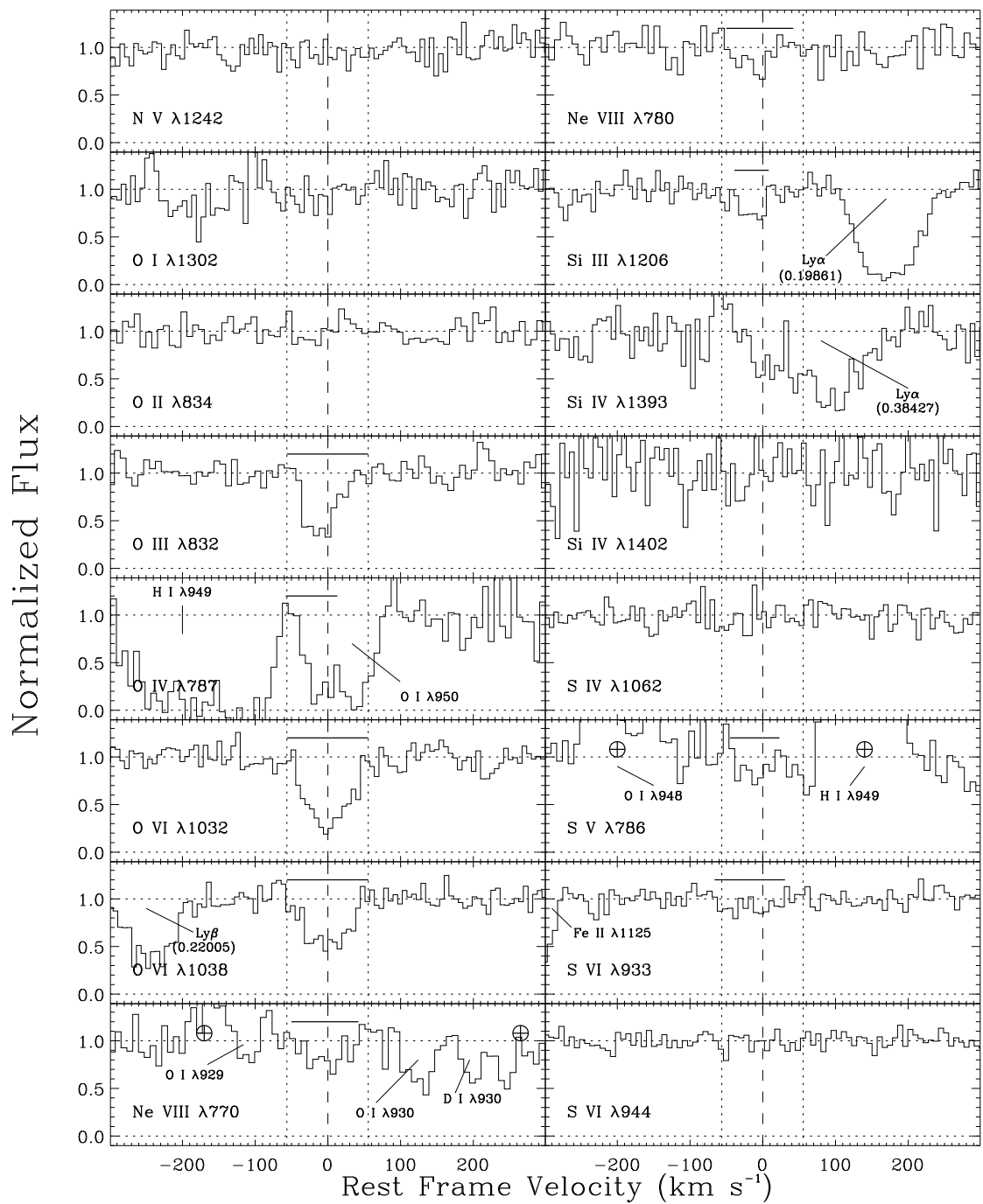

Fig. 2.— continued.

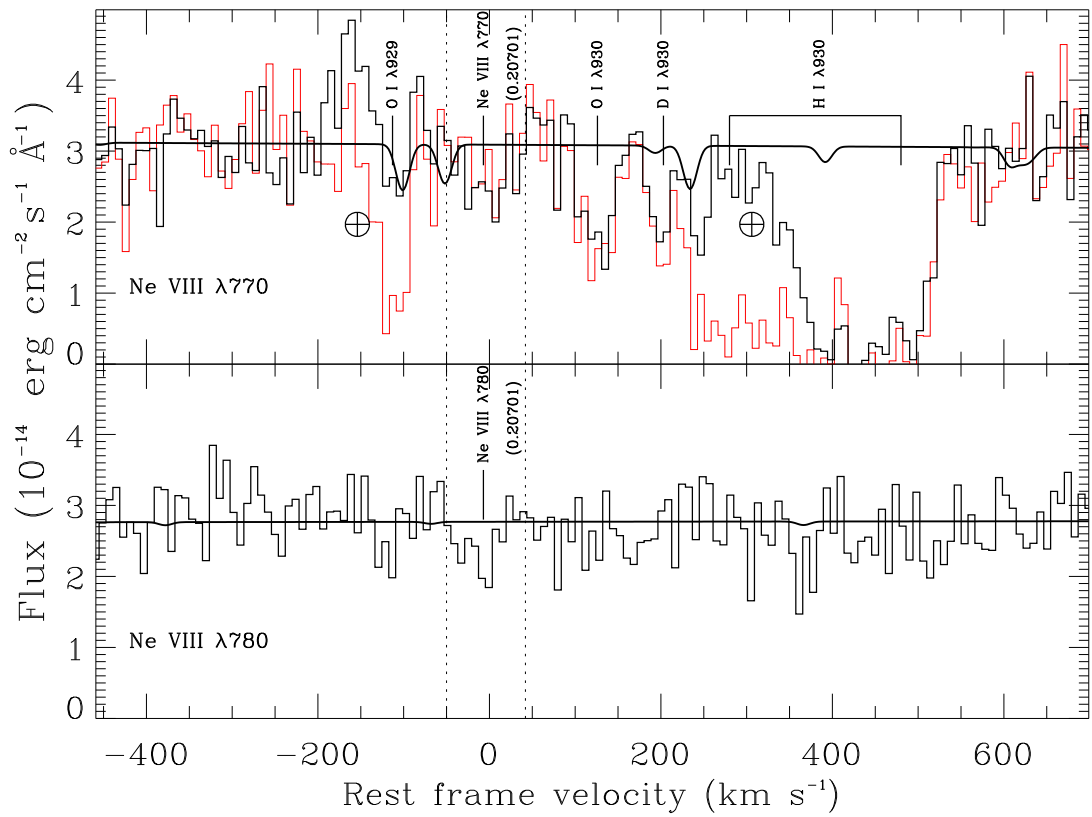

Fig. 3.—

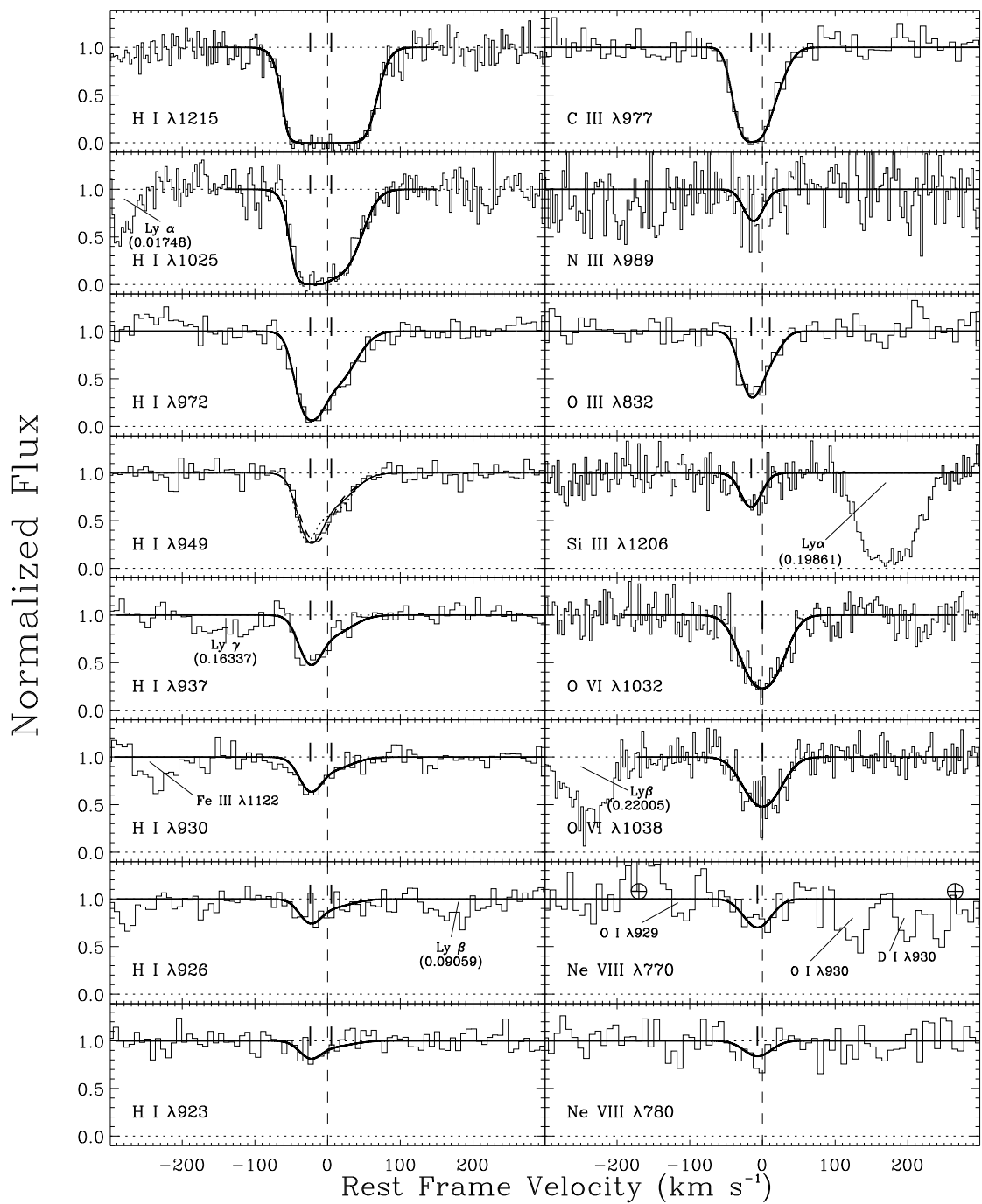

Fig. 4.—

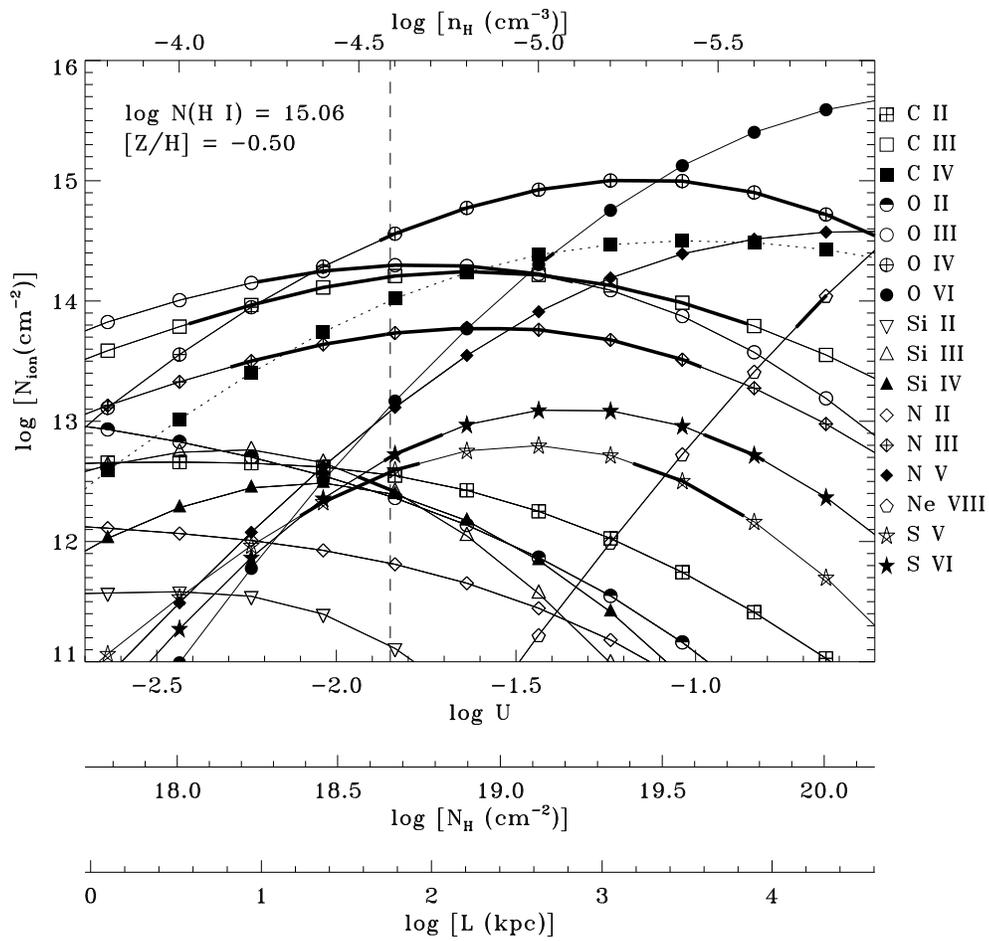

Fig. 5.—

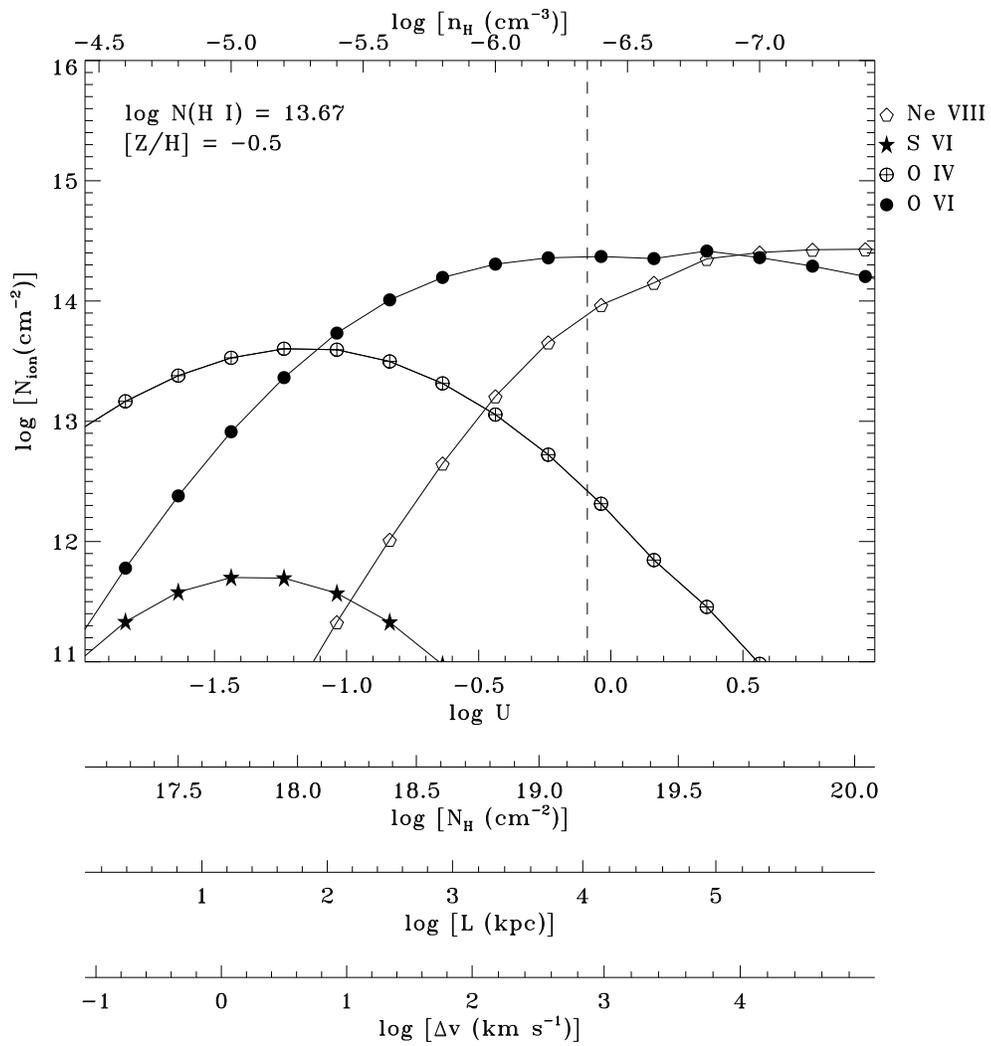

Fig. 6.—

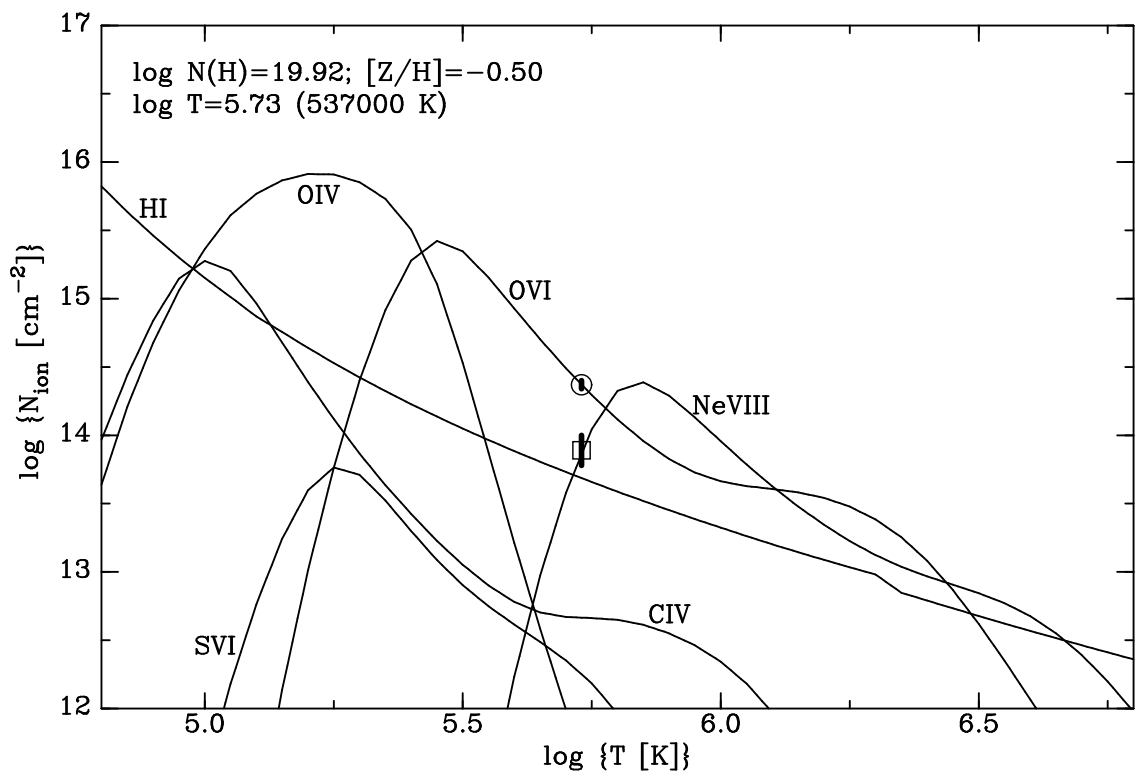

Fig. 7.—

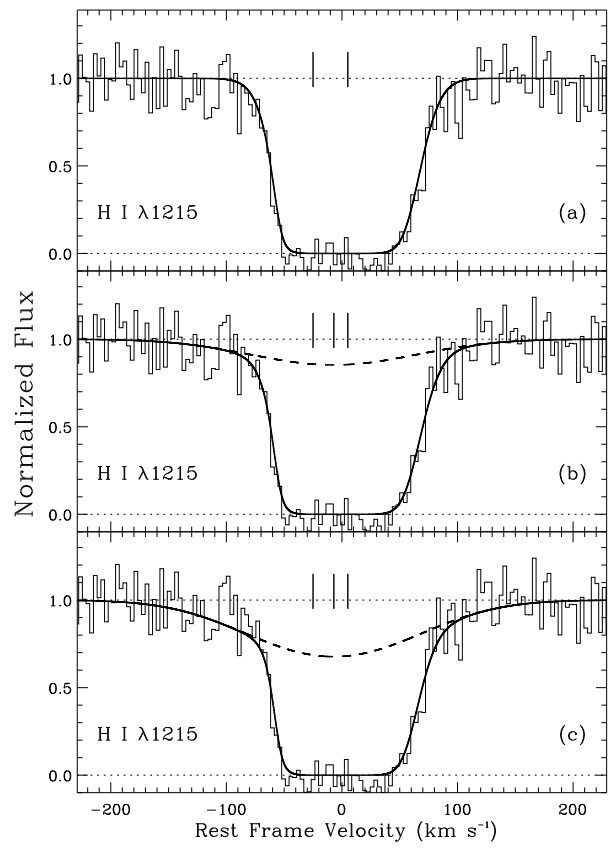

Fig. 8.—